\documentclass[10pt]{iopart}
\usepackage{iopams}  
\expandafter\let\csname equation*\endcsname\relax
\expandafter\let\csname endequation*\endcsname\relax
\usepackage{amsmath}
\usepackage{upgreek}
\usepackage{graphicx}
\usepackage{cite}
\usepackage{cancel}
\usepackage{cuted}
\usepackage{chngcntr}
\usepackage{xcolor}
\begin{document}
\title[Thermoelectric Properties of Type-I and Type-II Nodal Line Semimetals: A Comparative Study\ldots]
{Thermoelectric Properties of Type-I and Type-II Nodal Line Semimetals: A Comparative Study}
\author{Mohammad~Norman~Gaza~Laksono$^{1,2*}$, M~Aziz~Majidi$^{2}$, 
Ahmad~R.~T.~Nugraha$^{1,3}$}
\address{$^1$Research Center for Quantum Physics, 
National Research and Innovation Agency (BRIN), 
South Tangerang 15314, Indonesia}
\address{$^2$Department of Physics, 
Faculty of Mathematics and Natural Sciences, 
Universitas Indonesia, Depok 16424, Indonesia}
\address{$^3$ Department of Engineering Physics, 
Telkom University, Bandung 40257, Indonesia}
\address{$^{*}$Author to whom any correspondence should be addressed.}
\eads{\mailto{ahma080@brin.go.id \\ \hspace{28pt} mohammad.norman@ui.ac.id}}
\date{\today}
\begin{abstract}
We investigate the thermoelectric (TE) properties of nodal line semimetals (NLSs) using a combination of semi-analytical calculations within Boltzmann's linear transport theory and the relaxation time approximation, along with first-principles calculations for the so-called type-I and type-II NLSs. We consider the conduction and valence bands that cross near the Fermi level of these materials through first-principles calculations of typical type-I (TiS) and type-II (Mg$_3$Bi$_2$) NLSs and use the two-band model fit to find the Fermi velocity $v_{F}$ and effective mass $m$ that will be employed as the initial energy dispersion parameters.  The optimum curvature value for each energy band is searched by tuning both $v_{F}$ and $m$ to improve the TE properties of the NLSs. By systematically comparing all of our calculation results, we observe that tuning $v_{F}$ significantly improves TE properties in both types of NLS compared to tuning $m$.  We also find that in all TE metrics, the type-I NLS surprisingly can surpass the type-II NLS, which seems counter-intuitive to the fact that within the two-band model, the type-I NLS contains a parabolic band while the type-II NLS possesses a higher-order, Mexican-hat band. Our study demonstrates that optimizing the curvature of energy bands by tuning $v_F$ can significantly improve the TE performance of NLSs. This approach could guide future efforts in exploring other semimetals as potential TE materials by manipulating their band structures. 
\end{abstract}

\submitto{\PS}
\maketitle
\ioptwocol
\section{Introduction}

Among the primary energy sources such as gas, oil, and coal that are consumed by humans, it has been estimated that only one-third is used effectively, and two-thirds is wasted, most of which are in the form of heat~\cite{FITRIANI2016635,koumoto2013thermoelectric,elsheikh2014review, zheng2014review}. This form of energy can be converted into useful electrical energy through the so-called thermoelectric (TE) materials. Unfortunately, TE devices often have lower efficiency than most energy conversion schemes. One can assess the TE performance through some parameters such as figures of merit: $ZT = S^{2} \sigma T/\kappa$, where $S$ is the Seebeck coefficient, $\sigma$ is the electrical conductivity, $T$ is the operating temperature, and $\kappa$ is the total thermal conductivity. The thermal conductivity here is a sum of the electronic thermal conductivity ($\kappa_{e}$) and the lattice thermal conductivity ($\kappa_{ph}$), $\kappa = \kappa_{e}+\kappa_{ph}$. The figure of merit is also proportional to the power factor ($\mathrm{PF}$) by the following relation: $ZT = \mathrm{PF} \cdot T/\kappa$, where $\mathrm{PF}=S^{2}\sigma$. From this expression, we can see that a good TE material should possess good electrical conduction and thermal isolation. In other words, one should maximize the Seebeck coefficient and electrical conductivity to obtain a good TE material, while simultaneously, the thermal conductivity should be minimized. Unfortunately, $\sigma$ and $\kappa$ are strongly coupled with each other, hence making it difficult to find the material with high $ZT$~\cite{zeier2016thinking,yang2016tuning,jain2016computational,zhu2017compromise,he2017advances,gorai2017computationally,mao2018advances,darmawan2022thermoelectric,hu2016evidence}. The interplay among these parameters is primarily governed by the Wiedemann-Franz law, which states that the ratio between $\sigma$ and $\kappa_{e}$ is constant. Therefore, obtaining a material with high $\sigma$ while possessing low thermal conductivity $\kappa$ is very challenging since $\kappa$ also increases when $\sigma$ is enhanced.

Many attempts to decouple the interdependent TE parameters have been proposed to obtain as large $ZT$ as possible for various materials. \linebreak Some examples of such efforts are carrier concentration optimization \cite{CHEN2012535,Zhou2016SnTe}, nanostructuring materials \cite{poudel2008high,ma2008enhanced,xie2009unique,xie2010identifying,yan2013thermoelectric}, band convergence engineering \cite{mahan1996best,mao2015high}, \linebreak and hierarchical architecture consideration \cite{biswas2012high,ren2017enhancing}. Of various methods used to scan potential TE materials, the band engineering methods such as tuning the gap~\cite{Pei2011Ag2Te} and the effective mass \cite{bilc2004resonant, Heremans2012bulk, pei2012low} in terms of the curvature of the band could be effective because these methods use a relatively cheap computational method by considering only the energy dispersion relation $E(k)$ and the scattering lifetime $\uptau(E)$. By doing band-gap tuning or changing the combination of the band structure, one can obtain the optimized structure that will give better TE properties. Several works related to band engineering method have been performed for many types of band structures, such as pudding-mold bands \cite{usui2017enhanced, usui2013large, kuroki2007pudding, isaacs2019remarkable, wei2020strain}, parabolic bands \cite{xia2019leveraging,xia2019high, adhidewata2022thermoelectric}, and the linear Dirac bands \cite{adhidewata2022thermoelectrics,hasdeo2019optimal}.

Metals and semimetals are usually not considered as good TE materials due to their poor performance originating from the absence of the energy gap which makes the contributions of electrons and holes in the Seebeck coefficient cancel each other. By contrast, the existence of the heavy bands alongside the Dirac bands gives a high PF value in a semimetal like $\mathrm{CoSi}$~\cite{xia2019high}. Recently, materials with non-trivial band topology such as the nodal line semimetals (NLSs)~\cite{Shao2020, Rudenko2018,Singha2017, ali2016butterfly, Guan2017}, in which the conduction and valence bands intersect in the form of a line (called the nodal line), have received some attention due to their unique properties and characteristics. The NLSs can be classified into type-I and type-II NLSs based on the slopes of the bands along their nodal lines. The type-I NLS possesses two bands with oppositely aligned slopes near the nodal line and one of its bands is tilted slightly, while the type-II NLS has bands with the same slope near the nodal line because one of its bands is completely tipped over ~\cite{zhang2017topological}. Other works have also shown that some NLS phases found in $\mathrm{Nb}_{3}\mathrm{GeTe_{6}}$~\cite{wang2020unique} or $\mathrm{YbMnSb}_{2}$~\cite{pan2021thermoelectric} might be promising for TE applications with a Seebeck coefficient twice that of normal metals.

It should be noted that the existence of an intersection between a heavy band and Dirac bands at the Fermi level is found to enhance PF due to the improved electron-phonon scattering in the form of a sharp spike density of states (DOS) \cite{xia2019leveraging,rudderham2021ab}. Specifically, the existing heavy band acts as a filter for the low-energy carriers to be excited \cite{bahk2016minority, androulakis2010thermoelectric}. This unbalanced condition will lead to the Seebeck coefficient enhancement \cite{gayner2016boost}. However, the effects of specific electronic band properties such as band curvature and slope on TE properties of type-I NLS are yet unknown. Furthermore, we wonder that although the type-II NLS was claimed to be a promising TE material \cite{hung2022enhanced}, a systematic comparison of TE properties between type-I and type-II NLSs is not available.  Regarding this fact, we are fascinated to find out which type of NLSs will have the higher enhancement of TE performance.

In this work, we will discuss the TE properties of type-I (type-II) NLS materials by using $\mathrm{TiS}$ and $\mathrm{Mg}_{3}\mathrm{Bi}_{2}$ as model materials for each type, respectively. We calculate TE properties by employing a two-band model where we consider two energy bands of each NLSs, namely conduction and valence bands, near the Fermi level. Then, we will tune the value of energy dispersion parameters from each band which also alters the shape of the band itself, and see its implication on the TE properties of each NLSs.  In Section \ref{sec:th}, we show the model of the band structure and DOS that are considered for each type of NLSs.  The band parameters extracted from the first-principles band structures are used in the semi-analytical methods to obtain TE properties within the relaxation time approximation (RTA). Section \ref{sec:res} contains our results from the model; it consists of two subsections that discuss the TE properties of type-II and type-II NLSs, respectively. This paper is concluded in Section \ref{sec:con}.

\section{ Model and Methods}\label{sec:th}

In this section, we begin by outlining the band structure model for each type of NLS considering the energy dispersion of each band. Then, we will show how to apply this model to our semi-analytical calculations. Lastly, we give the computational parameters used in our first-principles calculations.

\subsection{Two-band model} \label{sec:2BM}

The TE properties can be calculated using the Boltzmann transport theory with RTA. In this approach, we express the TE properties (Seebeck coefficient, electrical conductivity, and electron thermal conductivity) in terms of the TE integrals $\mathcal{L}_{i}$ as \cite{goldsmid2010introduction,hasdeo2019optimal,wang2020unique,pan2021thermoelectric,ashcroft1976nd}

\begin{equation} \label{eq: Seebeck coefficient TE integral}
    S=\frac{1}{eT} \frac{\mathcal{L}_{1}}{\mathcal{L}_{0}}
\end{equation}

\begin{equation} \label{eq: Electrical conductivity TE integral}
    \sigma = e^{2} \mathcal{L}_{0} , 
\end{equation}
and 
\begin{equation} \label{eq: Thermal conductivity TE integral}
    \kappa_{e} = \frac{1}{T} \left(\mathcal{L}_{2}-\frac{\mathcal{L}_{1}^{2}}{\mathcal{L}_{0}} \right),
\end{equation}
respectively. Where $\mathcal{L}_{i}$ depends on the transport properties of the material according to
\begin{equation} \label{eq:transport-properties}
    \mathcal{L}_{i} = \int \uptau v^{2} g(E) \left(-\frac{\partial f}{\partial E}\right) (E-\mu)^{i} dE . 
\end{equation}
In Eq.~(\ref{eq:transport-properties}), $i={0,1,2}$, $\tau$ is the relaxation time, \break $v = \hbar^{-1} |\nabla_{\mathbf{k}}{E / \sqrt{3}}|$ is the electron longitudinal velocity, $g(E)$ is the density of states (DOS), and
$f(E)$ is the Fermi-Dirac distribution:
\begin{equation}
    f(E) = \frac{1}{1+\exp[(E-\mu)/k_{B}T]}
\end{equation} with its partial derivative with respect to energy $E$:
\begin{equation} 
\frac{\partial f}{\partial E} = -\frac{\exp \left[(E-\mu)/k_{B}T \right]}{(1+\exp[(E-\mu)/k_{B}T])^{2}} \, \left(\frac{1}{k_{B}T} \right)
\end{equation}
where $k_{B}$, $\mu$, and $T$ are Boltzmann constant, chemical potential, and temperature, respectively. We include $1/eT$ in Eq.~(\ref{eq: Seebeck coefficient TE integral}) since $S$ itself is defined as heat per carrier per temperature. In Eq.~(\ref{eq: Thermal conductivity TE integral}), $\kappa_e$ is defined as the amount of heat transferred through a material due to a temperature gradient which is why we include $1/T$. Eq.~(\ref{eq:transport-properties}) is often computed by integration of all bands available over the entire range of energy $E=[-\infty, \infty]$. Nevertheless, for a large number of materials, the thermoelectric characteristics depend mainly on the structure of the electronic bands near the Fermi level \cite{markov2018semi}. We also need to remember that Eq.~(\ref{eq:transport-properties}) is only defined in one band formulation. 

Since we consider two bands as the main contributors to the TE properties, we break down Eq.~(\ref{eq:transport-properties}) as the sum of conduction and valence band contributions, i.e., \cite{goldsmid2010introduction}


\begin{equation} \label{eq: conduction band contribution}
    \mathcal{L}_{c,i} = \int_{E_{0},c}^{\infty} \tau v^{2} g(E) \left(- \frac{\partial f}{\partial E} \right) (E-\mu)^{i} \, dE,
\end{equation} and

\begin{equation} \label{eq: valence band contribution}
    \mathcal{L}_{v,i} = \int_{- \infty}^{E_{0,v}} \tau v^{2} g(E) \left(- \frac{\partial f}{\partial E} \right) (E-\mu)^{i} \, dE
\end{equation} 
where $E_{0,c}$ and $E_{0,v}$ denote the energy at the band edge of the conduction band and of the valence band, respectively. The procedure is justified as we can see from the previous work \cite{chasapis2015understanding}. Following this division, the TE properties of our material can also be decomposed into conduction and valence band components, (denoted by $c$ and $v$ subscript respectively), such that the total TE transport coefficients are given by 

\begin{equation} \label{eq: Seebeck coefficient from total contribution}
    S=\frac{S_{c} \sigma_{c}+S_{v} \sigma_{v}}{\sigma_{c}+\sigma_{v}},
\end{equation}
\begin{equation} \label{eq: electrical conductivity from total contribution}
    \sigma = \sigma_{c} + \sigma_{v},
\end{equation}
and
\begin{equation} \label{eq: thermal conductivity from total contribution}
    \kappa_{e} = \frac{\sigma_{c} \sigma_{v}}{\sigma_{c}+\sigma_{v}} (S_{c}-S_{v})^{2} + (\kappa_{e,c}+ \kappa_{e,v}).
\end{equation} 
We will apply Eqs.~\eqref{eq: conduction band contribution}--\eqref{eq: thermal conductivity from total contribution} to our model of type-I and type-II NLSs. In our calculations, we do not consider the lattice contribution for the thermal conductivity since we would like to assess the TE performance of the NLSs in their optimistic scenario.

\subsubsection{Type-I NLS}

In the type-I NLS, the nodal line is formed when the two bands cross with different slope directions. In this model, we use a Dirac band as the conduction band, and a parabolic band as the valence band. The energy dispersion is given by \cite{neto2009electronic,ariel2012energy}
\begin{equation}\label{eq:Type-I NLS conduction band energy dispersion}
    E_{c}(k) = \hbar v_{F} \mathbf{|k|} , 
\end{equation}
for the conduction band and
\begin{equation}\label{eq:Type-I NLS valence band energy dispersion}
    E_{v}(k) = -\frac{\hbar^{2}{|\mathbf{k}|^{2}}}{2m} + E_{0}
\end{equation}
for the valence band, respectively, where $\mathbf{k}$ is the electron wavevector, $v_{F }$ is the Fermi velocity of the Dirac band, $m$ is the effective mass at the edge of the band, and $E_{0}$ an energy parameter that is used to determine the position of the valence band maximum. We also define $g(E)$ and $v(E)$ for conduction band as:

\begin{equation} \label{eq: DOS conduction band}
    g_{c}(E) = \frac{E^{2}}{\pi^{2}\hbar^{3} v_{F}^{3}},
\end{equation} and
\begin{equation} \label{eq: Longitudinal velocity conduction band}
    v_{c}(E) = \frac{v_{F}}{\sqrt{3}}
\end{equation} respectively, while for the valence band, we define it as
\begin{equation} \label{eq: DOS valence band}
    g_{v}(E) = \frac{m}{\hbar^{2} \pi^{2}} \sqrt{\frac{2m(E-E_{0})}{\hbar^{2}}},
\end{equation} and

\begin{equation} \label{eq: Longitudinal velocity valence band}
    v_{v}(E) = -\frac{1}{\sqrt{3}} \sqrt{\frac{-2(E-E_{0})}{m} .}
\end{equation}

By substituting the above definitions of $g(E)$ and $v(E)$ into Eqs.~(\ref{eq: conduction band contribution}) and~(\ref{eq: valence band contribution}), we obtain:

    \begin{equation}
        \begin{split} \label{eq: conduction band contribution_typeI}
            \mathcal{L}_{c,i} &= \frac{(k_{B}T)^{i+2}}{3 \pi^{2} \hbar^{3} v_{F}} \int_{0}^{\infty} \tau (x+\eta)^{2} \, x^{i} \frac{\exp(x)}{(1+\exp(x))^{2}} \, dx,
        \end{split}
    \end{equation}
and
    \begin{equation} \label{eq: valence band contribution_typeI} 
    \begin{split}
        \mathcal{L}_{v,i} =&  \frac{2(k_{B}T)^{i+3/2} \sqrt{2m}}{3\pi^{2}\hbar^{3}} \\
        & \times \int_{-\infty}^{\varepsilon_{0}} \tau (-x-\eta+\varepsilon_{0})^{3/2} \, 
         x^{i} \frac{\exp(x)}{(1+\exp(x))^{2}} \, dx. 
    \end{split}
    \end{equation}
    

Here, $\mu$ and $E_{0}$ parameters have been reduced with the thermal energy $k_{B}T$ such that $\eta = \mu/k_{B}T $ and $\varepsilon_{0}=E_{0}/k_{B}T$, respectively.

\subsubsection{Type-II NLS} 

\begin{figure*}
    \centering
    \includegraphics[width=14cm,clip]{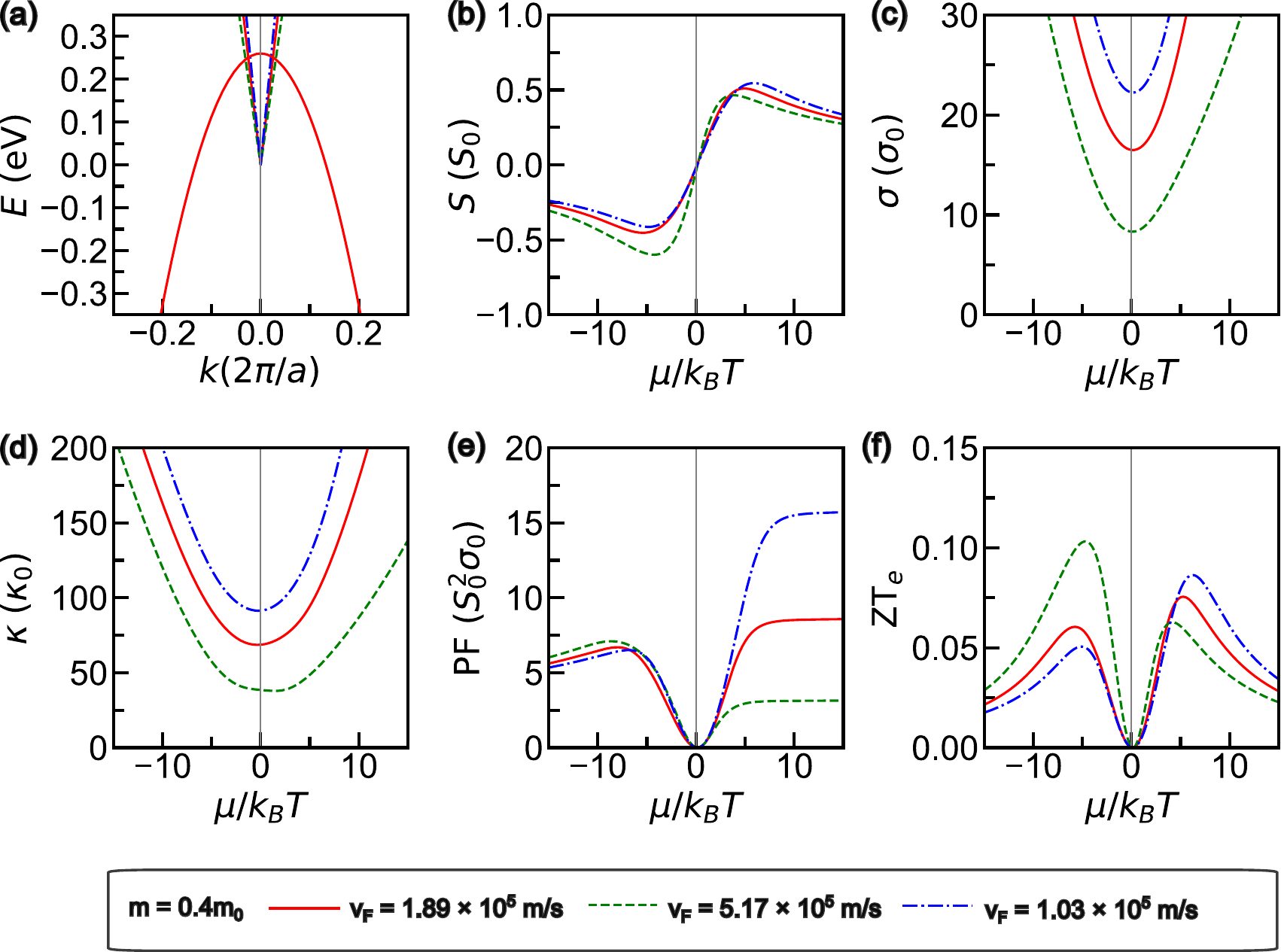}
    \caption{ Energy dispersion and TE properties of a type-I NLS model with a varying value of Fermi velocity $v_{F}$. For each combination of the two-band model, we show (a)~the energy dispersion, (b)~Seebeck coefficient $S$, (c)~electrical conductivity $\sigma$, (d)~electronic thermal conductivity $\kappa_{e}$, (e)~power factor (PF), and (f)~electronic figure of merit $ZT_e$. TE properties are plotted versus reduced chemical potential $\mu / k_{B}T$. The results for $S$, $\sigma$, and $\kappa_{e}$ are expressed in the units of $S_{0}$, $\sigma_{0}$, and $\kappa_{0}$ respectively.}
    \label{fig:Type-I velocity}
\end{figure*}

For a type-II NLS, the crossing between the valence and conduction bands occurs when their slopes have the same direction. We model this using the Dirac band for the conduction band and the Mexican-hat-shaped valence band. The energy dispersion of the valence band is given by \cite{wickramaratne2015electronic}:
\begin{equation} \label{eq:Mexican-hat band dispersion}
    E_{v}(k) = \frac{(\hbar^{2}k^{2}/(4m)-E_{1})^{2}}{E_{1}}+E_{0},
\end{equation}
where $E_{1}$ signifies the depth of the central valley of the central valley of the Mexican-hat band measured from the band edge, $m$ is defined at the valley in the middle of the Mexican-hat band. We define $v(E)$  and $g(E)$ for type-II NLS as:
\begin{equation}\label{eq:velocity type-II NLS}
    v_{\pm}(\varepsilon)=\sqrt{\frac{k_{B}T(\varepsilon_{0}-\varepsilon)}{3m}} \sqrt{1 \pm \sqrt{(\varepsilon_{0}-\varepsilon)/\varepsilon_{1}}} 
\end{equation} 
and
\begin{equation}\label{eq:DOS type-II NLS}
    g_{\pm}(\varepsilon)= \frac{4m}{\pi^{2}\hbar^{2}} \sqrt{\frac{m k_{B}T\varepsilon_{1}}{\hbar^{2}}} \sqrt{\frac{1 \pm \sqrt{(\varepsilon_{0}-\varepsilon)/\varepsilon_{1}}}{(\varepsilon_{0}-\varepsilon)/\varepsilon_{1}}} \:.
\end{equation} 

Note that since the Mexican-hat band has a valley in the middle of the band, there are two different values for the velocity, one for the outer ring (with a ``$+$" sign) and the other for the inner ring (with a ``$-$" sign). Substituting Eqs.~(\ref{eq:velocity type-II NLS}) and (\ref{eq:DOS type-II NLS}) into Eq.~(\ref{eq: valence band contribution}), we obtain the following equations:
\begin{equation} \label{eq: Lviout}
    \begin{split}
            \mathcal{L}^{out}_{v,i}  =& \frac{4(k_{B}T)^{i+3/2} \sqrt{m}}{3 \pi^{2} \hbar^{3}} \\
             & \times \int_{-\infty}^{\varepsilon_{0}-\eta}   \tau  \varepsilon_{1} \sqrt{\varepsilon_{0}-x-\eta}
             \left(1+\sqrt{(\varepsilon_{0}-x-\eta)/\varepsilon_{1}} \right)^{3/2} \\ 
             & \times  x^{i} \frac{\exp(x)}{(1+\exp(x))^{2}} \, dx. \\
    \end{split}
\end{equation}
for the outer ring and
\begin{equation} \label{eq: Lviin}
    \begin{split}
        \mathcal{L}^{in}_{v,i} =& \frac{4(k_{B}T)^{i+3/2} \sqrt{m}}{3 \pi^{2} \hbar^{3}} \\ 
        & \times \int_{\varepsilon_{0}-\varepsilon{1}-\eta}^{\varepsilon_{0}-\eta} \tau \varepsilon_{1} \sqrt{\varepsilon_{0}-x-\eta} \left(1-\sqrt{(\varepsilon_{0}-x-\eta)/\varepsilon_{1}} \right)^{3/2} \\ 
        & \times \frac{\exp(x)}{(1+\exp(x))^{2}} x^{i} \, dx \\
    \end{split}
\end{equation}
for the inner ring. We also scale $\mu$, $E_{0}$, and $E_{1}$ parameters with $k_{B}T$ as $\eta=\mu/k_{B}T$, $\varepsilon_{0}=E_{0}/k_{B}T$, and $\varepsilon_{1}=E_{1}/k_{B}T$, respectively.

We express the units of TE properties as $S_{0}$, $\sigma_{0}$, and $\kappa_{0}$.  The value of $S_0$ is the same for type-I and type-II NLSs, i.e.,  $S_{0}~=~k_{B}/e \approx~86.17~\upmu\mathrm{V/K}$.  On the other hand, we distinguish the values of $\sigma_0$ and $\kappa_0$ depending on the NLS types.  For the type-I NLS, we have 
\begin{equation} \label{eq:sigma0 type-I NLS}
    \sigma_{0} = \frac{2 \tau e^{2}(k_{B}T)}{3 \pi^{2} \hbar^{2}} \, \sqrt{\frac{2 m k_{B}T}{\hbar^{2}}}
\end{equation}
and
\begin{equation} \label{eq:kappa0 type-I NLS}
    \kappa_{0} = \frac{2 \tau k_{B}^{3}T^{2}}{3 \pi^{2} \hbar^{2}} \, \sqrt{\frac{2m k_{B}T}{\hbar^{2}}},
\end{equation}
while for the type-II NLS, we have
\begin{equation} \label{eq:sigma0 type-II NLS}
    \sigma_{0} =  \frac{4 \tau e^{2} (k_{B}T)}{3 \pi^{2}\hbar^{3}} \, \sqrt{\frac{ m k_{B}T}{\hbar^{2}}} 
\end{equation}
and
\begin{equation} \label{eq:kappa0 type-II NLS}
    \kappa_{0} = \frac{4 \tau k_{B}^{7/2} T^{5/2} \sqrt{m}}{3 \pi^{2} \hbar^{3}}.
\end{equation}

\subsection{First-principles simulations}
We perform first-principles calculations for both types of NLSs by using Quantum ESPRESSO \cite{giannozzi2009quantum} to obtain the electronic properties that will be used to calculate the TE properties and compare them using the model aforementioned above. In this work, the parameters for the TiS and Mg$_3$Bi$_2$ as the model materials for the type-I and type-II NLSs, respectively, are obtained from AFLOWLIB database \cite{curtarolo2012aflowlib}. For the exchange-correlation functional, we employ the generalized gradient approximation (GGA) \cite{perdew1992atoms} of the Perdew-Burke-Ernzerhof (PBE) functional. We set the cutoff energy to $300 \,\, \mathrm{eV}$, which is already sufficient for the convergence. We also calculate the TE properties of the materials using BoltzTraP2 \cite{madsen2018boltztrap2}, a package that works based on the Boltzmann transport equation to be compared with our model.  The electronic properties, i.e., band structures and DOS, from the first-principles calculation can be seen in \ref{sec:appendix1}.  We fit those band structures with the model energy dispersion for each energy band and tune its curvature through varying $v_{F}$ and $m$.

\section{Results and discussion}
\label{sec:res}

\begin{figure*}
    \centering
    \includegraphics[width=0.7\textwidth]{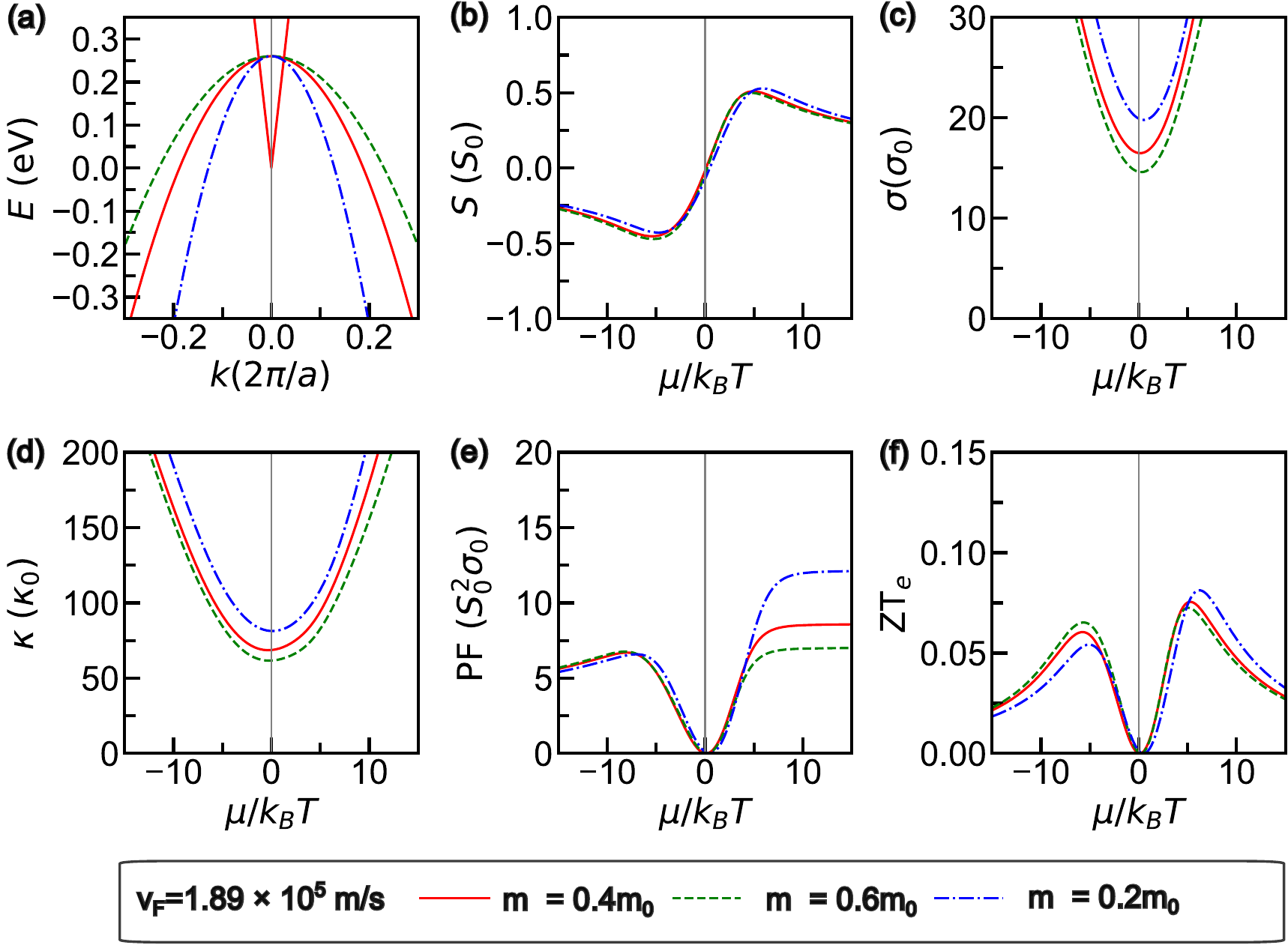}
    \caption{Energy dispersion and TE properties of a type-I NLS model with a varying value of hole effective mass $m$. For each combination of the two-band model, we show (a)~the energy dispersion, (b)~Seebeck coefficient $S$, (c)~electrical conductivity $\sigma$, (d)~electronic thermal conductivity $\kappa_{e}$, (e)~power factor (PF), and (f)~electronic figure of merit $ZT_{e}$. TE properties are plotted versus reduced chemical potential $\mu / k_{B}T$. The results for $S$, $\sigma$, and $\kappa_{e}$ are expressed in the units of $S_{0}$, $\sigma_{0}$, and $\kappa_{0}$ respectively.}
    \label{fig:Type-I mass}
\end{figure*}

Using Eqs.~(\ref{eq:Type-I NLS conduction band energy dispersion})--(\ref{eq:Type-I NLS valence band energy dispersion}), (\ref{eq: conduction band contribution_typeI})--(\ref{eq:Mexican-hat band dispersion}), and (\ref{eq: Lviout})--(\ref{eq: Lviin}), we will show the schematic plots of the energy dispersion and discuss the TE properties for both type-I and type-II NLSs. The energy dispersion plots are obtained by fitting the energy level coordinates from the first-principles calculations, from their respective reference materials $\mathrm{TiS}$ and $\mathrm{Mg}_{3}\mathrm{Bi}_{2}$ to our energy dispersion model equations (detailed in \ref{sec:appendix1}). This allows us to determine the energy dispersion parameters,  $v_{F}$ and $m$. 

From the fitting, we obtain $v_{F}~=~1.89~\times 10^{5}~\mathrm{m/s}$ and $m~=~0.4m_{0}$. These values are used in Eqs.~(\ref{eq:Type-I NLS conduction band energy dispersion})--(\ref{eq:Type-I NLS valence band energy dispersion}) and (\ref{eq:Mexican-hat band dispersion}) to plot the energy dispersion for each energy band, shown in red in panel (a) of Figs.~(\ref{fig:Type-I velocity})--(\ref{fig:Type-II mass}).  To investigate the effects of $v_{F}$ and $m$ on TE properties, we modify the shape of one of the energy bands by selecting higher or lower values of $v_{F}$ and $m$ than those initially obtained. Specifically, we set $v_{F}~=~5.17~\times 10^{5}~\mathrm{m/s}$ and  $v_{F}~=~1.03~\times ~10^{5}~\mathrm{m/s}$ for the Fermi velocity variation and $m~=~0.6m_{0}$ and $m~=~0.2m_{0}$  for the effective mass variation beyond those initial values obtained from the fitting.  Here, $m_{0}$ is the free electron rest mass in units of $\mathrm{MeV/c^{2}}$, i.e., $m_{0}~=  ~0.51099895~\,~ \mathrm{MeV}/\mathrm{c}^{2}  $.  The altered values of $v_F$ and $m$ are then used in our TE properties calculations, carried out semi-analytically using Eqs.~(\ref{eq: conduction band contribution_typeI})--(\ref{eq: valence band contribution_typeI}) and (\ref{eq: Lviout})--(\ref{eq: Lviin}) and with the help of SciPy package \cite{virtanen2020scipy} in Python for solving complicated integrals numerically. In all calculations of the TE properties, the temperature is fixed at $T = 300 \,\, \mathrm{K}$. 
The calculation codes are available in our GitHub repository \cite{GithubNorman}.  

The results for type-I NLS are shown in Figs.~\ref{fig:Type-I velocity}(a)--(f) and \ref{fig:Type-I mass}(a)--(f), while for type-II NLS are depicted in Figs.~\ref{fig:Type-II velocity}(a)--(f) and \ref{fig:Type-II mass}(a)--(f).  Note that in this work we assess the TE performance of the NLSs in their most optimistic scenario, i.e., when the contribution of $\kappa_{ph}$ to total $\kappa$ is neglected.  Therefore, the figure of merit $ZT$ is reduced to the electronic figure of merit $ZT_e$ in all of the results discussed in the following sections. 

\subsection{Type-I NLS} \label{subsec: Type-I NLS}

For the type-I NLS, we plot the energy dispersion and TE properties using the parameters mentioned earlier by varying $v_{F}$ and $m$ values in Figs.~\ref{fig:Type-I velocity}(a)-(f) and \ref{fig:Type-I mass}(a)--(f), respectively. We plot the TE properties as a function of a reduced dimensionless chemical potential $\mu/k_{B}T$ to observe the doping effect on TE properties by varying the position of the doping level which is described as how much electrons are reduced or added which will affect  $\mu/k_{B}T$ position. A negative $\mu/k_{B}T$ indicates that this is p-type doping where we move the doping level to the valence band by reducing electrons, while a positive $\mu/k_{B}T$ indicates n-type doping which means that the doping level is moved to the conduction band by adding electron. From Eqs.~(\ref{eq:sigma0 type-I NLS}) and (\ref{eq:kappa0 type-I NLS}), we obtain the conductivity units of our type-I NLS model as  $\sigma_{0} \approx 3.364 \times 10^{3} \,\, \mathrm{S/m}$ and $\kappa_{0} \approx 7.494 \times 10^{-3} \,\, \mathrm{W/m.k}$

\begin{figure*}
    \centering
    \includegraphics[width=0.7\textwidth]{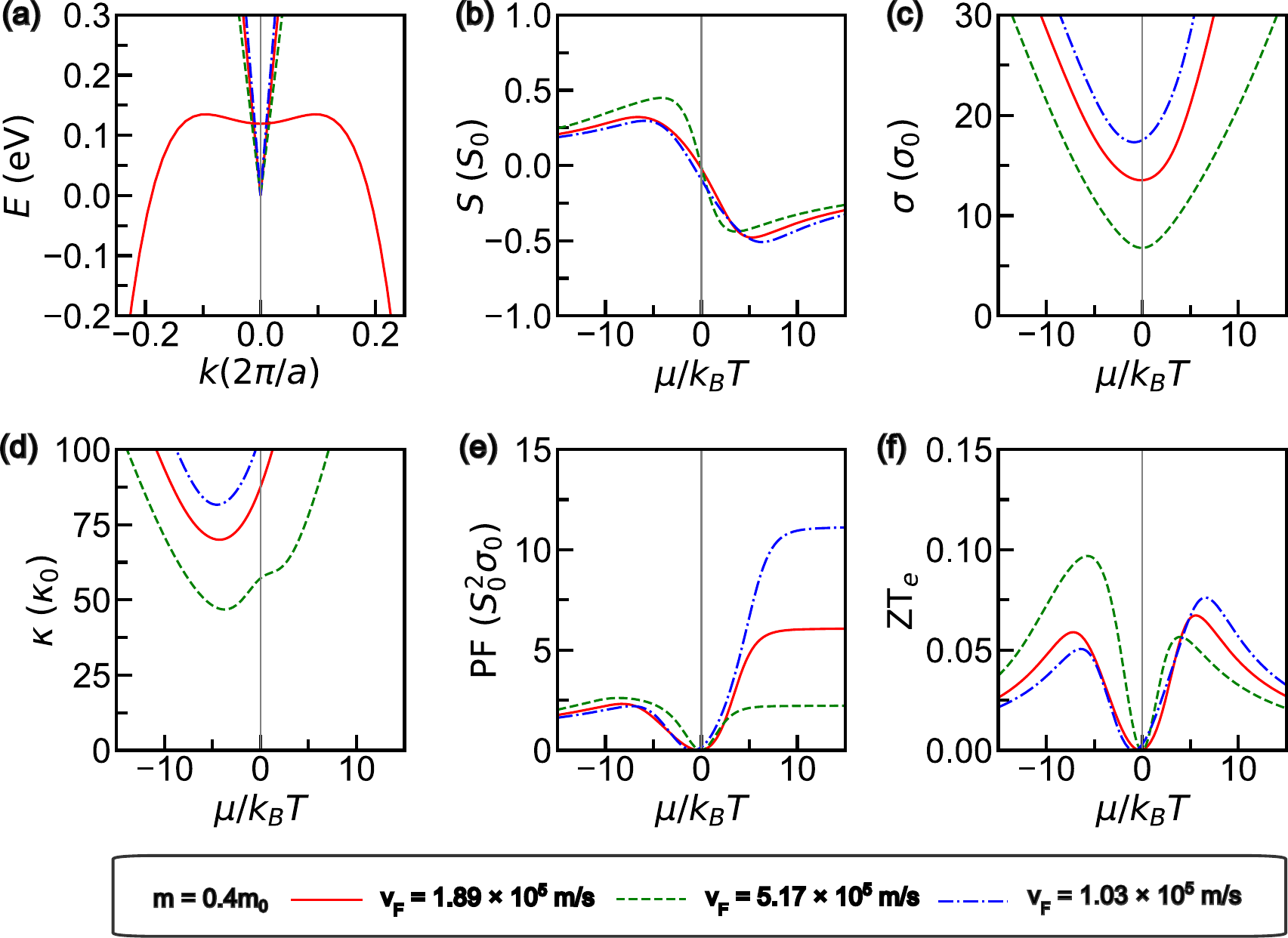}
    \caption{Energy dispersion and TE properties of a type-II NLS model with a varying value of Fermi velocity $v_{F}$. For each combination of the two-band model, we show (a)~the energy dispersion, (b)~Seebeck coefficient $S$, (c)~electrical conductivity $\sigma$, (d)~electronic thermal conductivity $\kappa_{e}$, (e)~power factor (PF), and (f)~electronic figure of merit $ZT_{e}$. TE properties are plotted versus reduced chemical potential $\mu / k_{B}T$. The results for $S$, $\sigma$, and $\kappa_{e}$ are expressed in the units of $S_{0}$, $\sigma_{0}$, and $\kappa_{0}$ respectively.}
    \label{fig:Type-II velocity}
\end{figure*}

In Figs.~\ref{fig:Type-I velocity}(a) and \ref{fig:Type-I mass}(a), we plot energy dispersion relations of each energy band of our type-I NLS model with varying values of $v_{F}$ and $m$, respectively. We observe that in Fig.~\ref{fig:Type-I velocity}(a) when $v_{F} = 1.89 \times 10^{5} \mathrm{m/s}$, a tiny increase on the slope of the Dirac band. On the other hand, when $v_{F}=1.03 \times 10^{5}$ the Dirac band becomes slightly steeper as shown in blue color in Fig.~\ref{fig:Type-I velocity}(a). Even though tuning the value of $v_{F}$ only makes a slight change in the slope of the Dirac band,  it does however greatly affect all TE properties values as we see in panels (b)--(f) of Fig.~\ref{fig:Type-I velocity}. Next, in Fig.~\ref{fig:Type-I mass} we see that changing $m$ greatly affects the shape of the parabolic band. When
$m = 0.2m_{0}$, the parabolic band becomes a light band, while when $m = 0.6m_{0}$, the parabolic band becomes a heavy band \cite{chasapis2015understanding}. Unfortunately, these great changes in the shape of the parabolic band do not affect TE properties compared to $v_F$ tuning. In the following part, we will consider how the changes in the shapes of these energy bands affect each of the TE properties.

First, we consider the Seebeck coefficient $S$ for our type-I NLS model in panel (b) of Figs.~\ref{fig:Type-I velocity} and~\ref{fig:Type-I mass} with varying $v_{F}$ and $m$ values, respectively. Both figures show an opposite trend in $|S|$ peak value changes in response to energy dispersion parameters tuning. We also observe a trend where $|S|$ peak value becomes negative when $\mu/k_{B}T$ is negative. The value is positive when $\mu/k_{B}T$ is shifted to the right where  $\mu/k_{B}T$ has positive values. The same trend is also observed in some materials that also have a parabolic band such as $\mathrm{KCaF}_{3}$ \cite{ali2019thermoelectric}. Initially, we set $v_{F} = 1.89 \times 10^{5} \mathrm{m/s}$ and $m = 0.4m_{0}$ in both figures as their initial values which are obtained from fitting. Then, we approximate the $|S|$ peak value from both figures using methods that are explained in \ref{sec:appendix2}. Based on our approximation results in Fig. \ref{fig:TypeI_output_vF}, we obtain $|S|$ peak value of $|0.509S_{0}|$ for n-type doping and $ |0.451S_{0}|$ for p-type doping.  We also observe that there is a significant shift towards $|S|$ when we tune the initial value of $v_{F}$. When we set $v_{F} = 5.17 \times 10^{5} \mathrm{m/s}$, the doping level shifts to the right, and $|S|$ peak value increases by $32.59\%$ for n-type doping in the left to $|0.598S_{0}|$ for the p-type doping which also marks the highest overall $|S|$ peak value we obtain for both type of NLSs. We also notice that the increase in $v_{F}$ greatly affects the $|S|$ peak value in p-type doping more than in n-type doping.  The only increase we obtain is when $v_{F}$ is set to $v_{F} = 1.03 \times 10^{5} \mathrm{m/s}$ where we only witness the $S$ peak value increases of $7.07\%$  which lower than $32.59\%$ increase that we obtain before in the p-type doping. 

\begin{figure*}
    \centering
    \includegraphics[width=0.7\textwidth]{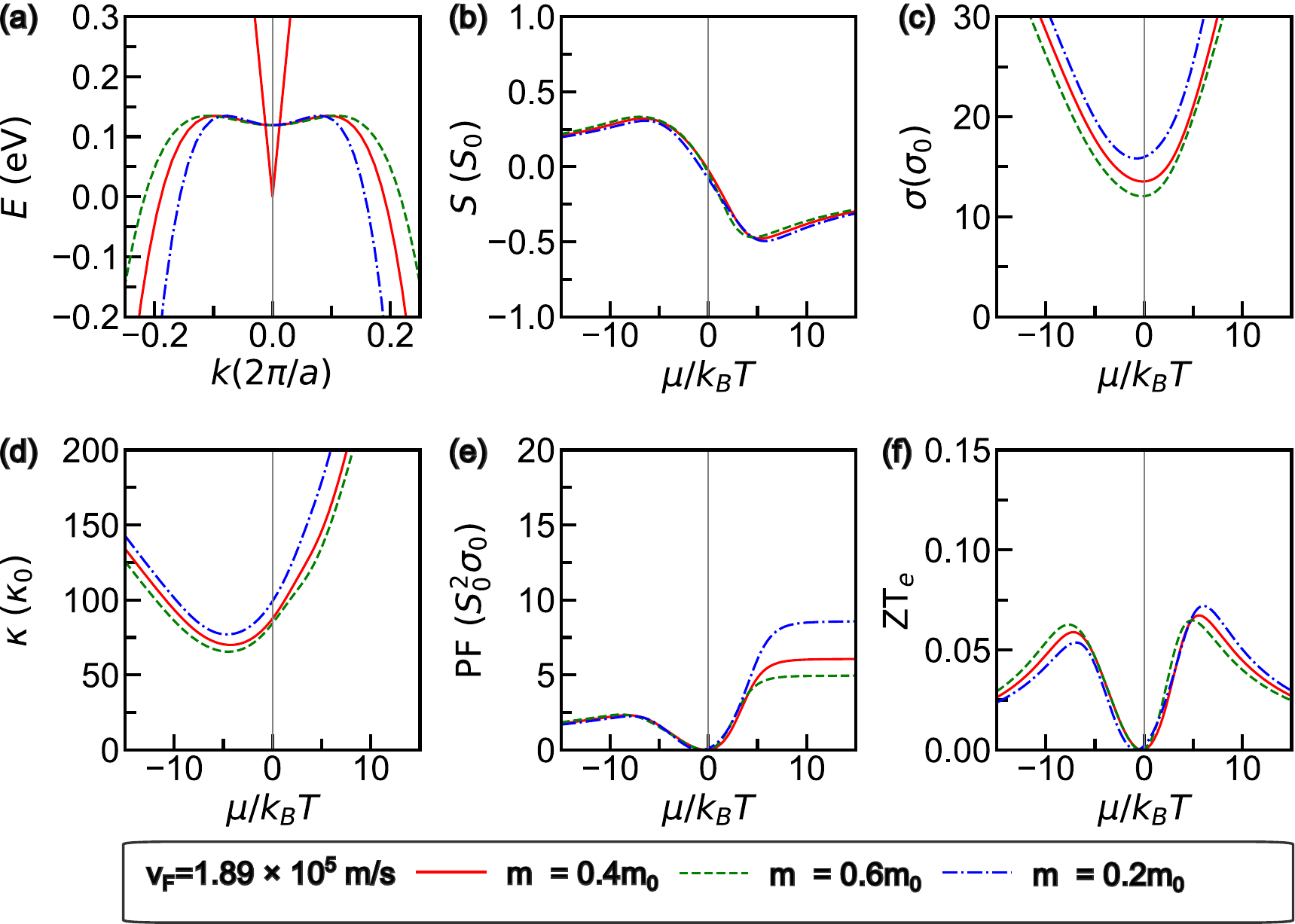}
    \caption{Energy dispersion and TE properties of a type-II NLS model with a varying value of hole effective mass $m$. For each combination of the two-band model, we show (a)~the energy dispersion, (b) ~Seebeck coefficient $S$, (c)~electrical conductivity $\sigma$, (d)~electronic thermal conductivity $\kappa_{e}$, (e)~power factor (PF), and (f)~electronic figure of merit $ZT_{e}$. TE properties are plotted versus reduced chemical potential $\mu / k_{B}T$. The results for $S$, $\sigma$, and $\kappa_{e}$ are expressed in the units of $S_{0}$, $\sigma_{0}$, and $\kappa_{0}$ respectively.}
    \label{fig:Type-II mass}
\end{figure*}

Next, apparently in Fig.~\ref{fig:Type-I mass}(b) tuning $m$ value does not have a significant change in $S$ peak value compared to tuning $v_{F}$. As we see from our approximation of $|S|$ peak value in Fig.~\ref{fig:TypeI_output_m}(a), we only obtain the highest increase in $|S|$ peak value when  $m = 0.2m_{0}$ where $|S|$ peak value increases only by $3.54\%$ to $|0.527S_{0}|$ for n-type doping. Additionally, we also observe that the doping level does not appear to be moving towards as it is only moved slightly to the left towards its initial position.

Furthermore, we consider the effect of tuning the shape of the energy bands on the magnitude of $\sigma$ and $\kappa$ in Figs.~\ref{fig:Type-I velocity}(c)--(d) and \ref{fig:Type-I mass}(c)--(d), respectively. Based on Fig.~\ref{fig:Type-I velocity}(c), we observe that both $\sigma$ and $\kappa$ increase as the Dirac band steepens, i.e. when $v_{F}~=~1.03~\times~10^5 ~\mathrm{m/s}$. Meanwhile, in ~\ref{fig:Type-I mass}(c), the increase occurs when the band gets narrower where this is obtained when $m~= ~0.2m_{0}$. In the same way, this is also true for $\mathrm{PF}$ and $ZT_{e}$ as can be seen Figs.~\ref{fig:Type-I velocity}(e)--(f) and \ref{fig:Type-I mass}(e)--(f), respectively. In Fig.~\ref{fig:TypeI_output_vF}(b), initially we obtain the PF value of $|8.56S_{0}^{2}\sigma_{0}|$ from initial $v_{F}$ value in the n-type doping and $|6.586S_{0}^{2}\sigma_{0}|$ in the p-type doping. We witness $83\%$ increase of PF peak value when we set $v_{F}~=~1.03~\times~10^5 ~\mathrm{m/s}$ in which PF peak value becomes $|15.67S_{0}^{2}\sigma_{0}|$ in the n-type doping. This value also marks the highest overall PF peak value from both types of NLSs. Unfortunately, in Fig.~\ref{fig:TypeI_output_m}(b) the highest PF peak value we obtain $41.18\%$ in the n-type doping where PF peak value is $|12.085S_{0}^{2}\sigma_{0}|$ also when $v_{F} = 1.03 \times 10^5 \mathrm{m/s}$.

Moreover, in all of our $ZT_{e}$ calculations we observe that we obtain higher value of nearby $\mu/k_{B}T = 0$ as can be seen in Figs.~\ref{fig:Type-I velocity}(f), \ref{fig:Type-I mass}(f), \ref{fig:Type-II velocity}(f), and \ref{fig:Type-II mass}(f). This enhancement can be attributed to our DOS calculation results in Figs.~\ref{fig:NLSs_electronic_properties}(b) and~\ref{fig:NLSs_electronic_properties}(d), which show a high DOS near the Fermi level denoted by the red dots in the Figure. The higher DOS values imply a greater number of available states for electrons to occupy, facilitating their excitation to higher energy levels. Consequently, these excited electrons exhibit increased mobility and contribute more effectively in electrical and thermal transport, resulting in higher $\sigma$ and $\kappa$ near the $\mu/k_{B}T = 0$, which also correlates with $ZT_{e}$. From Fig.~\ref{fig:TypeI_output_vF}(c) we initially obtain the $ZT_{e}$ value of $0.075$ for the n-type doping and $0.060$ for the p-type doping when $v_{F} = 1.89 \times 10^{5}~\,~\mathrm{m/s}$ for $ZT_{e}$ and $m = 0.4m_{0}$ in Fig.~\ref{fig:Type-I velocity}(f) and Fig.~\ref{fig:Type-I mass}(f). Then, we obtain $71.67\%$ increase of $ZT_{e}$ to $0.103$ when  $v_{F} = 1.03 \times 10^5 \mathrm{m/s}$ in the p-type doping in Fig.~\ref{fig:Type-I velocity}(f). This increase is the highest overall from both types of NLSs and also the highest $ZT_{e}$ value compared to type-II NLS. On the other hand, in Fig.~\ref{fig:Type-I mass}(f) where we tune the $m$ we only obtain $8.33\%$ increase of $ZT_{e}$ value to $0.065$ as can be seen in Fig.~\ref{fig:TypeI_output_m}(c).

We suspect that those increases aforementioned above result from the term $\frac{(k_{B}T)^{i+2}}{3 \pi^{2} \hbar^{3} v_{F}} $ in Eq.~(\ref{eq: conduction band contribution_typeI}) and the term $\frac{2(k_{B}T)^{i+3/2}\sqrt{2m}}{3\pi^{2}\hbar^{3}}$ in Eq.~(\ref{eq: valence band contribution_typeI}) for the Dirac band and parabolic band, respectively. This increase also agrees with the Wiedemann-Franz law.

\subsection{Type-II NLS} \label{subsec: Type-II NLS}

We plot the energy dispersion and TE properties for type-II NLS by varying the value of $v_{F}$ and $m$ in Figs.~\ref{fig:Type-II velocity}(a)–(f) and \ref{fig:Type-II mass}(a)–(f), respectively. For TE properties calculation we obtain from Eqs.~(\ref{eq:sigma0 type-II NLS}) and (\ref{eq:kappa0 type-II NLS})  $\sigma_{0}~\approx 4.757~\times 10^{3}~\mathrm{S/m}$ and $\kappa_{0}~\approx~5.299 ~\times 10^{-3}~\mathrm{W/m.k}$.

From Figs.~\ref{fig:Type-II velocity}(a) and \ref{fig:Type-II mass}(a), we observe that the given values of $v_{F}$ and $m$ greatly affect the slope of the Dirac band and the depth of the Mexican-hat band NLS type-II. For the Dirac band, it can be seen in Fig.~\ref{fig:Type-II velocity}(a) that the slope of the Dirac band will increase when $v_{F} = 5.17 \times 10^{5} \mathrm{m/s}$. On the other hand, when $v_{F} = 1.03 \times 10^{5} \mathrm{m/s}$, the Dirac band will get steeper and narrower. For the Mexican-hat band in Fig.~\ref{fig:Type-II mass}(a), the effect of the value of $m$ on the depth of the Mexican-hat band is as follows. As $m = 0.6m_{0}$, the depth of this band will increase. Contrarily, if $m = 0.2m_{0}$, the Mexican hat band will lose its depth. In the following part, we consider the effects of these energy band adjustments on type-II NLS TE properties.

First, we consider $S$ in  Figs.~\ref{fig:Type-II velocity}(b) and~\ref{fig:Type-II mass}(b). Initially, in Fig.~\ref{fig:TypeII_output_vF}(a) by the initial dispersion energy parameters, we obtain $|S|$ peak value of $|0.478S_{0}|$ for n-type doping and $|0.319S_{0}|$ for p-type doping, respectively. We obtain the highest $|S|$ peak value increase of about $40.75\%$ to $|0.449S_{0}|$ when $v_{F} = 5.17 \times 10^{5}~\mathrm{m/s}$ for the p-type doping. This positive value may be attributed to the inclination of the Dirac band. Accordingly, it is better to increase $v_{F}$ since it will also increase carrier mobility because the $S$ is also related to carrier mobility. In contrast, it also seems that tuning $m$ doesn't improve $|S|$ peak value significantly compared to tuning $v_{F}$ as seen in Fig.~\ref{fig:TypeII_output_m}(a). The highest increase we obtain only $3.56\%$ to $|0.495S_{0}|$ when $m=0.2m_{0}$ for the n-type doping. Therefore, we believe that tuning the shape of the Mexican hat band through $m$ tuning is not a good idea to improve the TE properties of our type-II NLS model compared to tuning the slope of the Dirac band.

Next, in Figs.~\ref{fig:Type-II velocity}(c)--(d) and \ref{fig:Type-II mass}(c)--(d) we can see that tuning $v_{F}$ and $m$ also cause changes in $\sigma$ and $\kappa$. Even though the increase is not as drastic as in the type-I NLS model, it still greatly affects PF and $\mathrm{ZT_{e}}$ as we explain in the later part. This increase is probably caused by the term $\frac{(k_{B}T)^{i+2}}{3 \pi^{2} \hbar^{3} v_{F}} $,
$\frac{4 \tau (k_{B}T)^{i+3/2}\sqrt{m}}{3 \pi^{2}\hbar^{3}}$, and $\frac{4 \tau (k_{B}T)^{i+3/2}\sqrt{m}}{3 \pi^{2} \hbar^{3}}$ in Eqs.~(\ref{eq: conduction band contribution_typeI}), (\ref{eq: Lviin}), and (\ref{eq: Lviout}), respectively.

Furthermore, in Figs.~\ref{fig:Type-II velocity}(e) and \ref{fig:Type-II mass}(e) where we consider the PF we obtain the respective PF of $|6.055S_{0}^{2}\sigma_{0}|$ for n-type doping and $|2.312S_{0}^{2}\sigma_{0}|$ for p-type doping using the initial energy dispersion parameters as can be seen in Figs.~\ref{fig:TypeII_output_vF}(b) and \ref{fig:Type-II mass}(b). The increase of PF peak value that occurs in Fig.~\ref{fig:TypeII_output_vF}(e) is about $83\%$ larger than the initial value when $v_F=1.03 \times 10^{5} \mathrm{m/s}$ which changes PF peak value to $|11.085S^{2}_{0}\sigma_{0}|$ for n-type doping. We also observe changes in PF peak value caused by $m$ tuning, although not as large as when we tune the $v_{F}$. We obtain a PF peak value increase of $41.29\%$ to $|8.555S_{0}^{2}\sigma_{0}|$ when $m=0.2m_{0}$, while for the p-type doping the increase we obtain is not significant. We observe that those increases in PF might be caused by a simultaneous increase in $\sigma$ in panel (c) of Figs.~\ref{fig:Type-II velocity} and \ref{fig:Type-II mass} through the relation $\mathrm{PF}=S^{2} \sigma$. Therefore, we suggest that tuning $v_{F}$ is the best option to obtain the optimum value for PF.

Finally, in Figs.~\ref{fig:Type-II velocity}(f) and \ref{fig:Type-II mass}(f) we obtain the respective $ZT_{e}$ peak values of  $0.067$ for n-type doping and $0,059$ p-type doping from our approximation results in Figs.~\ref{fig:TypeII_output_vF}(c) and \ref{fig:TypeII_output_m}(c) using the initial energy dispersion parameters, respectively. Additionally, according to our approximations of $ZT_{e}$ using another $v_{F}$ values, we obtain the highest increase in peak value of $ZT_{e}$ in Fig.~\ref{fig:Type-II velocity}(f) for the p-type doping when $v_{F} = 5.17 \times 10^{5}~\,~ \mathrm{m/s}$ where we obtain  $64.41\%$ increases in peak $ZT_{e}$ value to $0.097$. Meanwhile, we only obtain the highest $ZT_{e}$ increase of $7.46\%$ in Fig.~\ref{fig:Type-II mass}(f) when $m=0.2m_{0}$ based on our approximation done in Fig.~\ref{fig:TypeII_output_m}(c), which is also lower compared to our result in Fig.~\ref{fig:Type-I mass}(f). The highest increase of $ZT_{e}$ in Fig.~\ref{fig:Type-II velocity}(f) is smaller compared to Fig.~\ref{fig:Type-I velocity}(f) is because Eq.~(\ref{eq:kappa0 type-II NLS}) yields a higher $\kappa_{0}$ value compared to Eq.~(\ref{eq:kappa0 type-I NLS}) since it has $k_{B}$ and $T$ with higher order compared to Eq.~(\ref{eq:kappa0 type-I NLS}), which in turn very crucial in the calculation of $ZT_{e}$ because $\kappa_{0}$ act as the denominator. This also applies to Fig.~\ref{fig:Type-II mass}(f) and also the term $\sqrt{m}$ also affects our $ZT_{e}$ calculation because it also increases as $m$ increases.

In summary, those obtained TE properties values from both types of NLSs especially $S$ have values close to those of typical carbon-based TE composites at the same room temperature \cite{soleimani2020review}. Additionally, based on the results of $\sigma$ and $\kappa$ of both NLSs, these materials might also be considered for wearable TE generators (TEGs) since both possess high $\sigma$ values and have $\kappa$ that can be tuned through our methods of band engineering.

\section{Conclusions}
\label{sec:con}
We have systematically studied TE properties of NLS type-I and type-II through consideration of energy dispersion shapes by tuning each of the NLS band curvatures.  We found that changing the shape of the energy dispersion through slight modifications of its parameters, $v_{F}$ and $m$, can drastically affect the TE properties of both types of NLSs.  For the Seebeck coefficient $S$, we obtained the largest $S$ peak value in the p-doped type-I NLS of about $|0.598S_{0}|$, where $S_{0} \approx 86.17 \,\, \upmu \mathrm{V/K}$.  On the other hand, although the largest $|S|$ in the type-II NLS is less than that in the type-I NLS, the largest increase in $|S|$ up to $40.75\%$ is obtained by varying $v_F$ in the p-doped type-II NLS.  As for the power factor (PF), we observed about $83\%$ increase in the PFs of both types of NLSs in the n-doped regime, with the type-I NLS achieving the highest overall PF peak value compared to the type-II NLS.  Lastly, we recorded a $71.67\%$ increase in $ZT_e$ for the type-I NLS and a $64.41\%$ increase for the type-II NLS under p-type doping. From these results, we show that NLSs might be a potential TE material since these materials have TE properties close to some typical carbon-based TE composites at room temperature and probably would be promising in wearable TEGs application. Our work is expected to trigger further calculations to scan other potential TE materials, particularly in the class of semimetals, by manipulating their band structure through the variation of the curvatures of their energy bands. 

\section*{Acknowledgments}

 We acknowledge Mahameru BRIN for its HPC facility. M.N.G.L is supported by the research assistantship from the National Talent Management System at BRIN.

\appendix

\counterwithin{figure}{section}
\renewcommand{\thefigure}{A.\arabic{figure}}

\begin{figure}[tb!]
    \centering
    \includegraphics[width=0.9\linewidth]{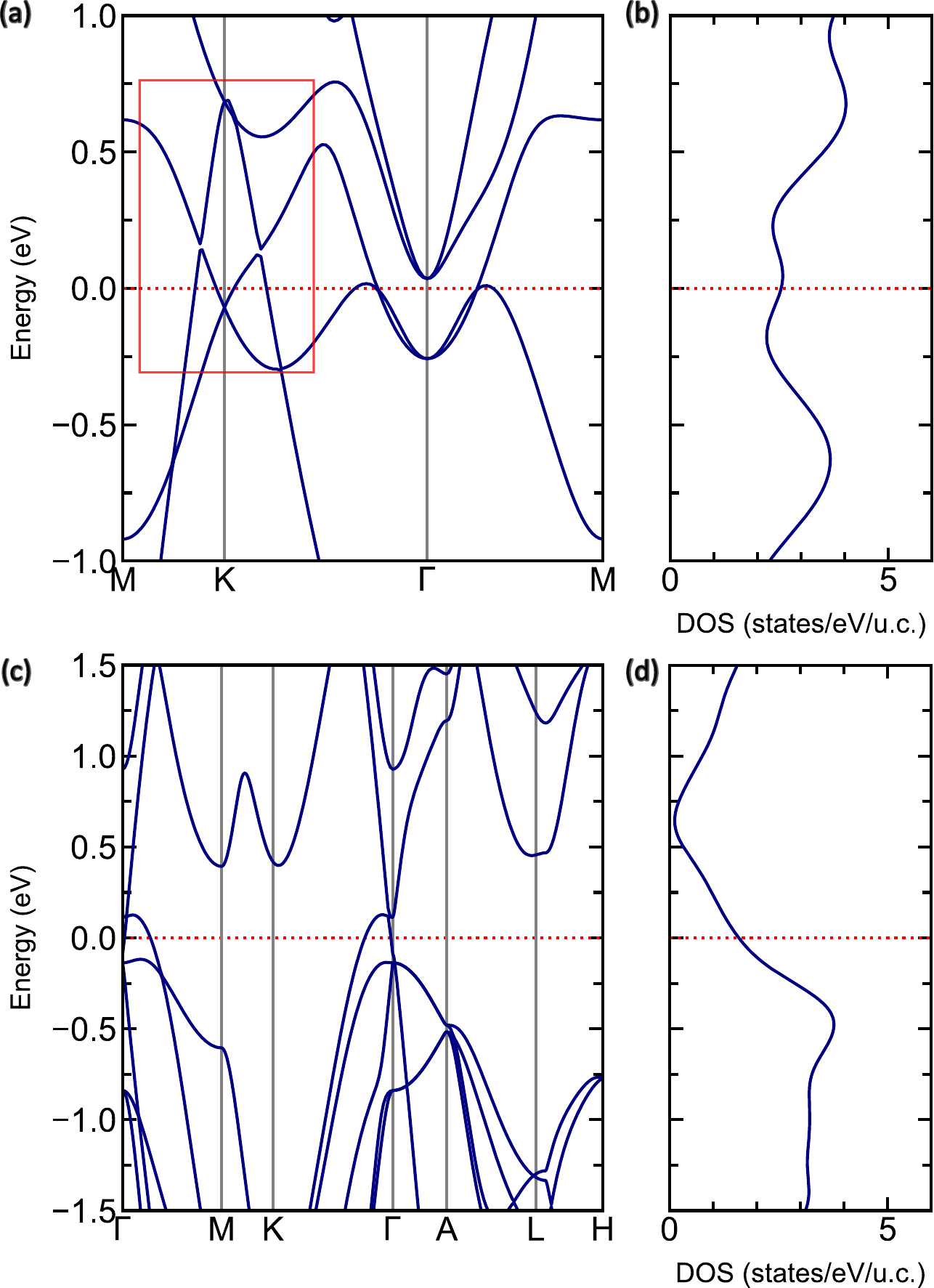}
        \caption{Electronic properties of NLS reference materials considered in this study: (a-b) $\mathrm{TiS}$ and (c-d) $\mathrm{Mg}_3\mathrm{Bi}_2$. Panels (a) and (c) show the results for the band structures, while panels (b) and (d) show the results for DOS. The Fermi level is shown in red dots line}
    \label{fig:NLSs_electronic_properties}
\end{figure}

\section{Electronic properties of TiS and Mg$_3$Bi$_2$ from first-principles simulations}
\label{sec:appendix1}

We use $\mathrm{TiS}$ and $\mathrm{Mg}_{3}\mathrm{Bi}_{2}$ as our reference materials for type-I and type-II NLS, respectively.  In Figs.~\ref{fig:NLSs_electronic_properties}(a)--(b), we show the electronic properties of $\mathrm{TiS}$.  From Fig.~\ref{fig:NLSs_electronic_properties}(a), we see a band crossing in $\mathrm{M}-\mathrm{K}-\Gamma$ path that confirms that the material belongs to type-I NLS. This also confirms the observation in Ref.~\cite{xu2020centrosymmetric}. Then, we take the energy band coordinates to fit with the dispersion model and tune its curvature.  Next, for $\mathrm{Mg}_{3}\mathrm{Bi}_{2}$, we show the electronic properties in Figs.~\ref{fig:NLSs_electronic_properties}(c)--(d).   From Fig.~\ref{fig:NLSs_electronic_properties}, we can see that there is a crossing in $\mathrm{K}-\Gamma-\mathrm{M}$ which confirms that the material is a type-II NLS, consistent with Ref.~\cite{zhang2017topological}. We perform a two-band model fitting for the type-II NLS similar to the case of type-I NLS to obtain the energy dispersion parameters, as mentioned at the beginning of Sec~\ref{sec:res}.

\counterwithin{figure}{section}
\renewcommand{\thefigure}{B.\arabic{figure}}

\section{Peak value approximations for the TE properties}\label{sec:appendix2}

Here we evaluate the TE properties by approximating their peak values to observe the change in the peak values of the TE properties due to variations of the energy dispersion parameters ($m$ and $v_{F}$). The peak values are approximated by numerically interpolating the TE properties using the NumPy package. 

\begin{figure}[tb!]
    \centering
    \includegraphics[width=\linewidth]{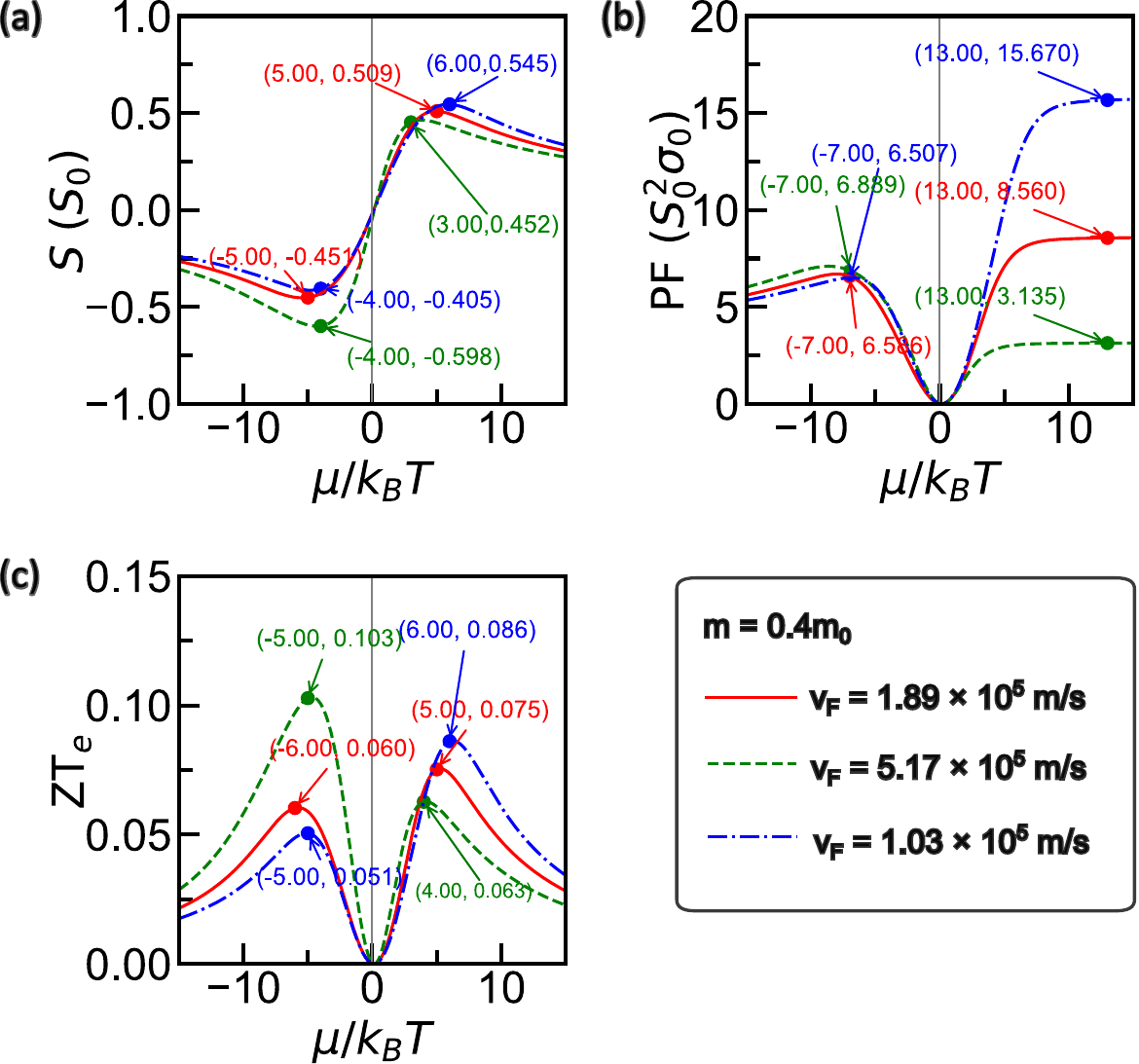}
    \caption{Approximations of peak values of TE properties for type-I NLS model with varying Fermi velocity $v_{F}$. We show the approximate peak values for (a) Seebeck coefficient $S$, (b) power factor (PF), and (c) electronic figure of merit $ZT_{e}$. }
    \label{fig:TypeI_output_vF}
\end{figure}

\begin{figure}[tb!]
    \centering
    \includegraphics[width=\linewidth]{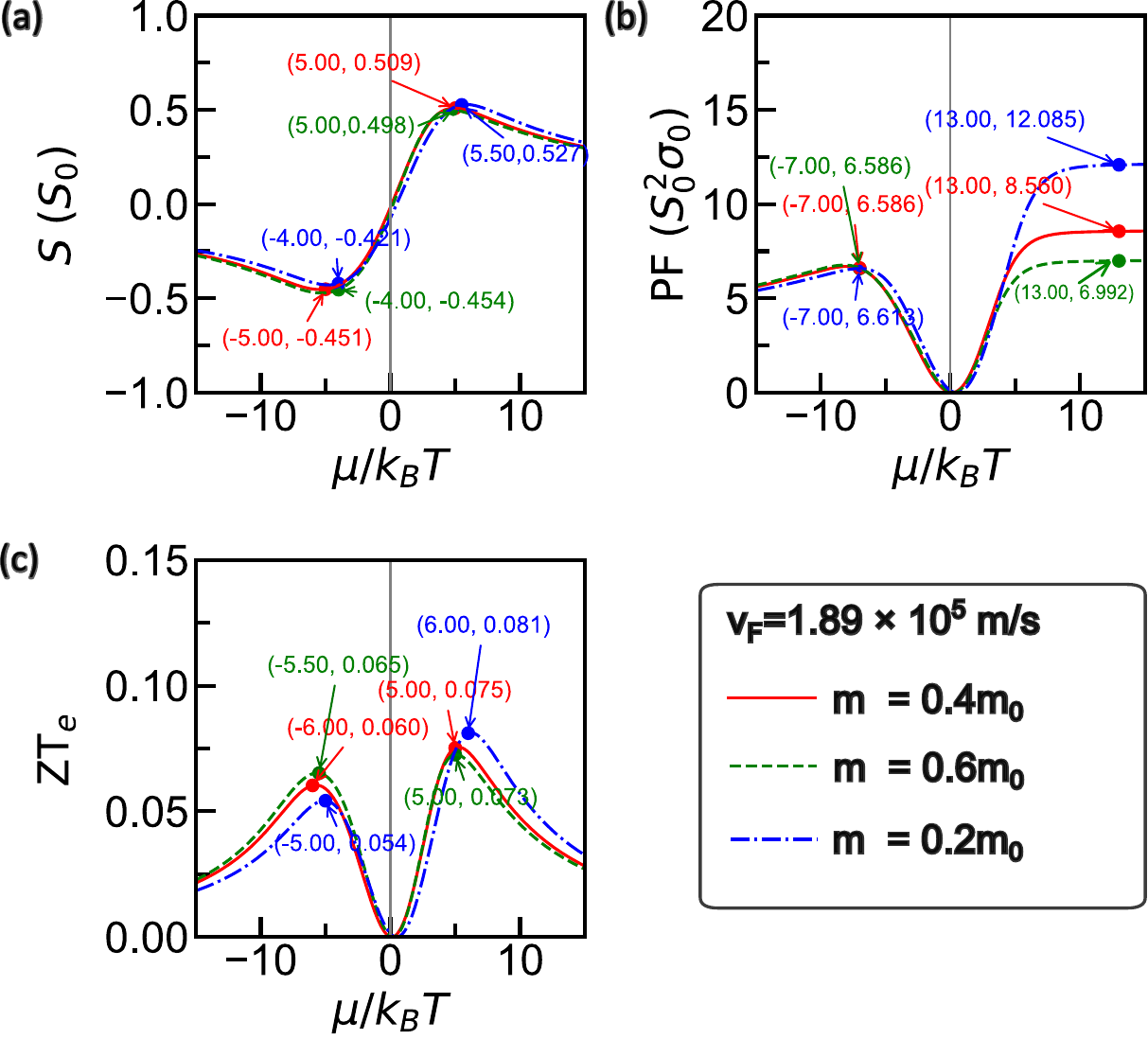}
    \caption{Approximations of peak values of TE properties for type-I NLS model with varying hole effective mass $m$ values. We show the approximate peak values for (a)~Seebeck coefficient $S$, (b)~power factor (PF), and (c)~electronic figure of merit $ZT_{e}$. }
    \label{fig:TypeI_output_m}
\end{figure}

In Figs.~\ref{fig:TypeI_output_vF}(a)--(c), we approximate peak values of the TE properties of the type-I NLS model by tuning the Dirac band with varying $v_{F}$ values.  For $S$ in
Fig.~\ref{fig:TypeI_output_vF}(a), we find that the initial $|S|$ peak values at $v_{F}~=~1.89~ \times~10^{5}~\mathrm{m/s}$ are $|0.509 S_{0}|$ for the n-type doping and $|0.451 S_{0}|$ the for p-type doping. As $v_{F}$ increases to  $v_{F}~=~ 5.17~\times~10^{5}~\mathrm{m/s}$, the $S$ peak value in the n-type doping decrease by $11.20\%$ to  $|0.452S_{0}|$, while for the p-type doping increases by $32.59\%$ to $|0.598S_{0}|$. Conversely, reducing $v_{F}$ to $v_{F}~=~1.03~\times~10^{5}~\mathrm{m/s}$ leads to a $7.07\%$ increase in the $|S|$ peak value for the n-type doping ($|0.545S_{0}|$) and a $10.20\%$ decrease for the p-type doping ($|-0.405S_{0}|$). These results indicate that modifying the Dirac band shape by tuning $v_{F}$ can significantly enhance the $|S|$ peak value in p-type doping when $v_{F}$ is increased while reducing $v_{F}$ results in a notable decrease.

In Fig.~\ref{fig:TypeI_output_vF}(b), we observe a different trend for the PF peak values. At $v_{F} = 5.17\times 10^{5}~\mathrm{m/s}$, the PF peak value for the n-type doping decreases sharply by $63.38\%$ to $|3.135S_{0}^{2}\sigma_{0}|$, while for the p-type doping, it increases slightly by $4.6\%$ to $|6.689_{0}^{2}\sigma_{0}|$. On the other hand, when $v_{F}$ is reduced to $v_{F}~=~1.03~\times~10^{5}~\mathrm{m/s}$, the PF peak value for the n-type doping increases dramatically by $83\%$ to $|15.67S_{0}^{2}\sigma_{0}|$, whereas for the p-type doping, it decreases modestly by $1.2\%$ to $|6.507S_{0}^{2}\sigma_{0}|$. Therefore, tuning $v_{F}$ significantly impacts the PF peak value in the n-type doping than in the p-type doping.

Finally, for $ZT_{e}$ in Fig.~\ref{fig:TypeI_output_vF}(c),  we observe the highest overall fluctuations in peak values, with both significant increases and decreases, in the type-I NLS model. The highest increase in $ZT_{e}$ occurs at $v_{F} = 5.17\times 10^{5} \mathrm{m/s}$ where it rises by $71.67\%$, reaching a peak value of $0.103$ which is also the highest $ZT_{e}$ peak value overall for type-I NLS. Then, the most notable decrease in $ZT_{e}$ is observed at $v_{F}~=~ 1.03~\times~10^{5}~\mathrm{m/s}$ where it drops by $15\%$ down to $0.051$.

\begin{figure}[tb!]
    \centering
    \includegraphics[width=\linewidth]{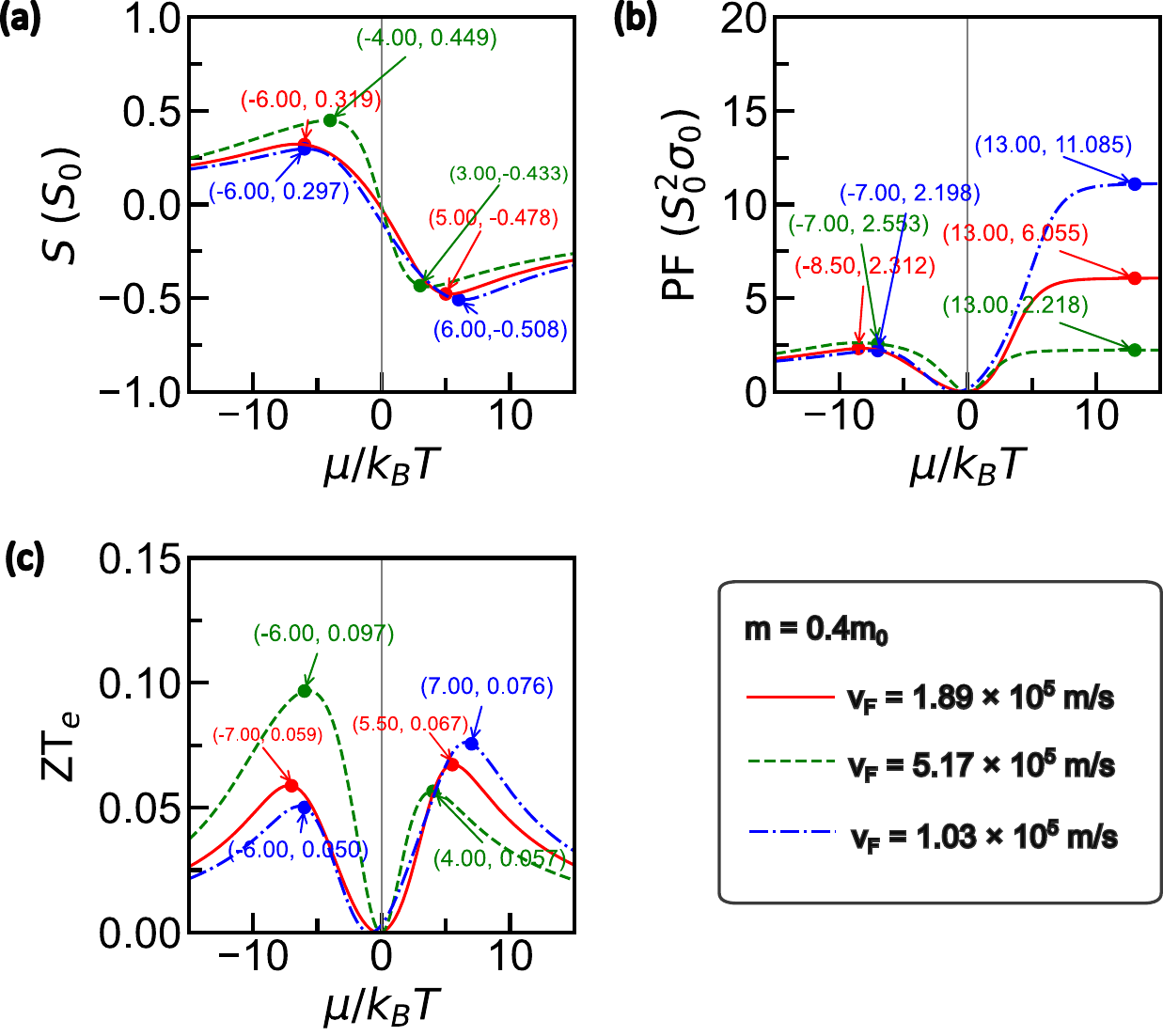}
    \caption{Approximations of peak values of TE properties for type-II NLS model with varying value of Fermi velocity $v_{F}$. We show the approximate peak values for (a)~Seebeck coefficient $S$, (b)~power factor (PF), and (c)~electronic figure of merit $ZT_{e}$. }
    \label{fig:TypeII_output_vF}
\end{figure}

Next, we analyze the TE properties peak value of the type-I NLS model by tuning the parabolic band with varying $m$ values in Figs.~\ref{fig:TypeI_output_m}(a)--(c). From Fig.~\ref{fig:TypeI_output_m}(a), the $|S|$ peak value decreases slightly by $2.16\%$ at  $m~= ~0.6m_{0}$ for the n-type doping, while for the p-type doping, it increases marginally by  $0.67\%$, reaching $|0.454S_{0}|$. At $m~=~0.2m_{0}$, the $|S|$ peak values show more noticeable changes, with an increase of $3.54\%$ for the n-type doping to $|0.527S_{0}|$, and a decrease of $6.65\%$ for the p-type doping to $|0.421S_{0}|$. These results suggest that tuning $m$ has a relatively minor effect on the type-I NLS TE properties, with only small changes observed in the peak values.

In Fig.~\ref{fig:TypeI_output_m}(b), we see an interesting trend for the PF peak value. At $m~= ~0.6m_{0}$, the PF peak value for the n-type doping decreases by $18.32\%$ to $|6.992S_{0}^{2}\sigma_{0}|$, while for the p-type doping, it increases about $0.41\%$ to $|6.613_{0}^{2}\sigma_{0}|$. However, when $m~=~0.2m_{0}$, the PF peak value for the n-type doping increases $41.18\%$ to $|12.085S_{0}^{2}\sigma_{0}|$, whereas for the p-type doping remains at $|6.586S_{0}^{2}\sigma_{0}|$.  These results highlight that tuning $m$ primarily only improves the PF in n-type doping.  Finally, for $ZT_{e}$ in Fig.~\ref{fig:TypeI_output_m}(c), the highest increase can be obtained by tuning $m$ is $8.33\%$ at $m~= ~0.6m_{0}$, where $ZT_{e}$ increases to $0.065$ for the p-type doping.  The highest $ZT_{e}$ peak value is obtained at $m~= ~0.2m_{0}$, where it increases by $8\%$ to $0.081$ in the n-type doping.

In Figs.~\ref{fig:TypeII_output_vF}(a)--(c), we approximate the peak values of TE properties of the type-II NLS model by tuning the $v_{F}$. As we see in Fig.~\ref{fig:TypeII_output_vF}(a), when $v_{F}~=~ 5.17~\times~10^{5}~\mathrm{m/s}$, the $S$ peak value in the n-type doping decreases by $9.41\%$ to  $|0.433S_{0}|$, while for the p-type doping it increases by $40.75\%$ to $|0.449S_{0}|$. Contrarily, as $v_{F}~=~1.03~\times~10^{5}~\mathrm{m/s}$, the $|S|$ peak value increases by $6.28\%$ for the n-type doping ($|0.508S_{0}|$). This $|S|$ increase is the largest and also the highest $S$ peak value overall for the type-II NLS.

\begin{figure}[tb!]
    \centering
    \includegraphics[width=\linewidth]{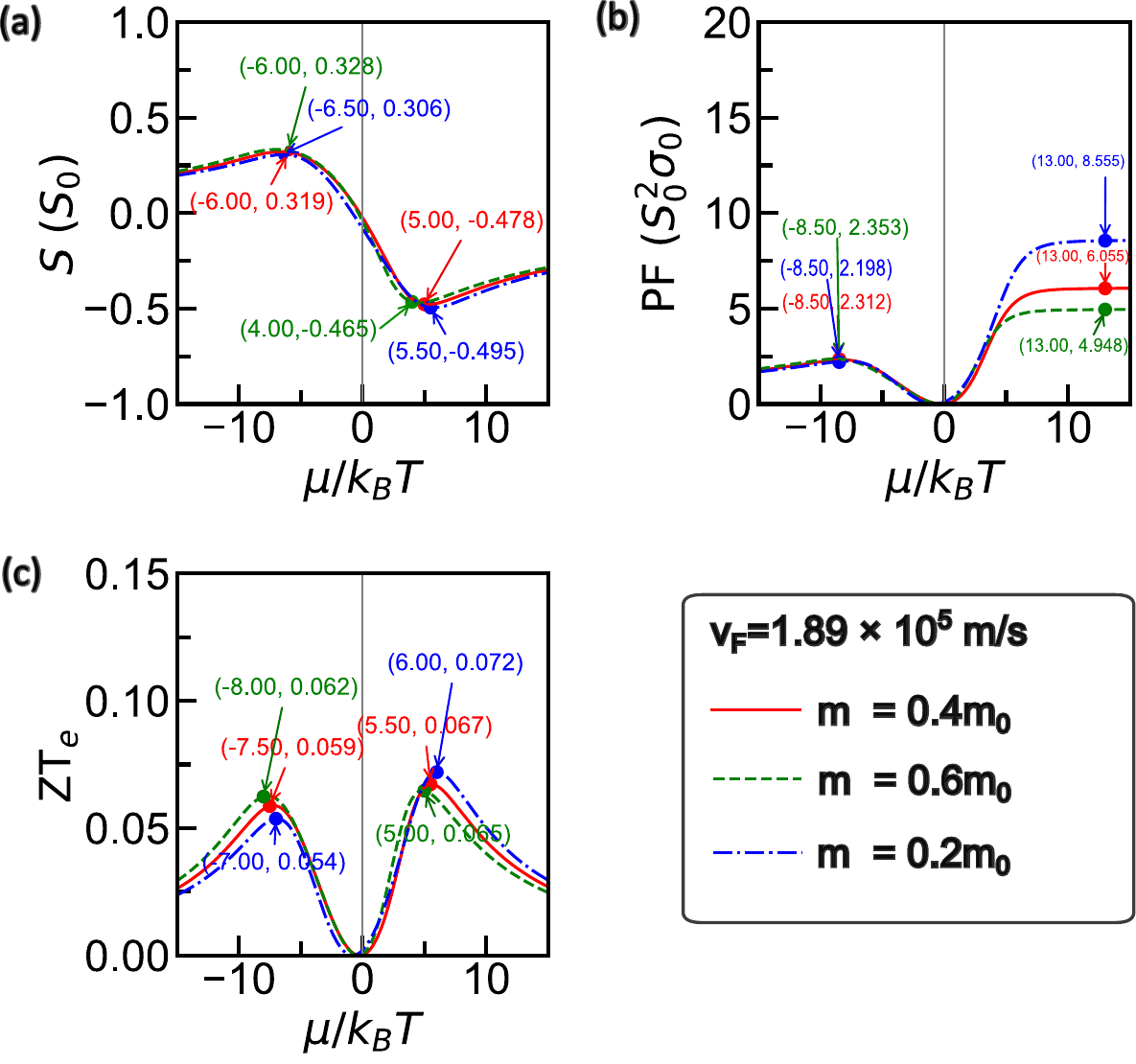}
    \caption{Approximations of peak values of TE properties for type-II NLS model with varying hole effective mass $m$ values. We show the approximate peak values for (a)~Seebeck coefficient $S$, (b)~power factor (PF), and (c)~electronic figure of merit $ZT_{e}$. }
    \label{fig:TypeII_output_m}
\end{figure}

Furthermore, from Fig.~\ref{fig:TypeII_output_vF}(b) we observe the PF peak value trend for type-II NLS when $v_{F}$ is tuned and also obtain that at $v_{F} = 1.03\times 10^{5}~\mathrm{m/s}$ the largest PF peak value for type-II NLS is $|11.085S_{0}^{2}\sigma_{0}|$ in the n-type doping which means it increases about $83\%$ relative to the value obtained using the initial energy dispersion parameter. In the type-II NLS, we also notice the same trend as that in the type-I NLS for the PF peak value that tends to be larger in the n-type doping rather than in the p-type doping. The PF peak values in the p-type doping also do not change much when we tune the energy dispersion parameters.  Then, based on Fig.~\ref{fig:TypeII_output_vF}(c),  we observe the highest overall changes in $ZT_{e}$ for the type-II NLS at $v_{F} = 5.17\times 10^{5} \mathrm{m/s}$, where it rises by $64.41\%$ to $0.097$ for the n-type doping. We also see that the changes in $ZT_{e}$ are more significant in the p-type doping than the n-type doping.

In the following, we examine the peak values of the TE properties for the type-II NLS model by tuning the Mexican-hat band with varying $m$ values in Figs.~\ref{fig:TypeII_output_m}(a)--(c). Based on Fig.~\ref{fig:TypeII_output_m}(a), we obtain the highest increase in the $|S|$ peak value by tuning $m$ for type-II NLS of about $3.56\%$ when $m~=~0.2m_{0}$ which changes the $|S|$ peak value to $|0.495S_{0}|$ for the n-type doping.  In Fig.~\ref{fig:TypeII_output_m}(b), for the PF peak value we obtain the highest increase of $41.29\%$ by tuning $m$, which makes the PF peak value becomes $8.555S_{0}^{2}\sigma_{0}$ in the n-type doping when $m~=~0.2m_{0}$. On the other hand, for the p-type doping, the change is only $1.77\%$ when $m~=~0.2m_{0}$, an this variation only leads to the PF peak value of about $2.353S_{0}\sigma_{0}^{2}$. Finally, in Fig.~\ref{fig:TypeII_output_m}(c), the highest increase of $ZT_{e}$ we obtain for type-II NLS by tuning $m$ is only $7.46\%$ at $m~= ~0.2m_{0}$ where $ZT_{e}$ increases to $0.072$ for the n-type doping.

\break
\section*{References}

\begin{thebibliography}{10}
\expandafter\ifx\csname url\endcsname\relax
  \def\url#1{{\tt #1}}\fi
\expandafter\ifx\csname urlprefix\endcsname\relax\def\urlprefix{URL }\fi
\providecommand{\eprint}[2][]{\url{#2}}

\bibitem{FITRIANI2016635}
Fitriani, Ovik R, Long B, Barma M, Riaz M, Sabri M, Said S and Saidur R 2016 {\em Renew. Sustain. Energy Rev.\/} {\bf 64} 635--659

\bibitem{koumoto2013thermoelectric}
Koumoto K and Mori T 2013 {\em Thermoelectric nanomaterials\/} (Berlin, Heidelberg: Springer)

\bibitem{elsheikh2014review}
Elsheikh M~H, Shnawah D~A, Sabri M~F~M, Said S~B~M, Hassan M~H, Bashir M~B~A and Mohamad M 2014 {\em Renew. Sustain. Energy Rev.\/} {\bf 30} 337--355

\bibitem{zheng2014review}
Zheng X, Liu C, Yan Y and Wang Q 2014 {\em Renew. Sustain. Energy Rev.\/} {\bf 32} 486--503

\bibitem{zeier2016thinking}
Zeier W~G, Zevalkink A, Gibbs Z~M, Hautier G, Kanatzidis M~G and Snyder G~J 2016 {\em Angew. Chem.\/} {\bf 55} 6826--6841

\bibitem{yang2016tuning}
Yang J, Xi L, Qiu W, Wu L, Shi X, Chen L, Yang J, Zhang W, Uher C and Singh D~J 2016 {\em npj Comput. Mater.\/} {\bf 2} 1--17

\bibitem{jain2016computational}
Jain A, Shin Y and Persson K~A 2016 {\em Nat. Rev. Mater.\/} {\bf 1} 1--13

\bibitem{zhu2017compromise}
Zhu T, Liu Y, Fu C, Heremans J~P, Snyder J~G and Zhao X 2017 {\em Adv. Mater.\/} {\bf 29} 1605884

\bibitem{he2017advances}
He J and Tritt T~M 2017 {\em Science\/} {\bf 357} eaak9997

\bibitem{gorai2017computationally}
Gorai P, Stevanovi{\'c} V and Toberer E~S 2017 {\em Nat. Rev, Mater.\/} {\bf 2} 1--16

\bibitem{mao2018advances}
Mao J, Liu Z, Zhou J, Zhu H, Zhang Q, Chen G and Ren Z 2018 {\em Adv. Phys\/} {\bf 67} 69--147

\bibitem{darmawan2022thermoelectric}
Darmawan A, Suprayoga E, AlShaikhi A~A and Nugraha A~R 2022 {\em Mater. Today Commun\/} {\bf 33} 104596

\bibitem{hu2016evidence}
Hu J, Tang Z, Liu J, Liu X, Zhu Y, Graf D, Myhro K, Tran S, Lau C~N, Wei J {\em et~al.\/} 2016 {\em Phys. Rev. Lett.\/} {\bf 117} 016602

\bibitem{CHEN2012535}
Chen Z~G, Han G, Yang L, Cheng L and Zou J 2012 {\em Prog. Nat. Sci.: Mater. Int.\/} {\bf 22} 535--549

\bibitem{Zhou2016SnTe}
Zhou M, Gibbs Z~M, Wang H, Han Y, Li L and Snyder G~J 2016 {\em Appl. Phys. Lett.\/} {\bf 109} 042102

\bibitem{poudel2008high}
Poudel B, Hao Q, Ma Y, Lan Y, Minnich A, Yu B, Yan X, Wang D, Muto A, Vashaee D {\em et~al.\/} 2008 {\em science\/} {\bf 320} 634--638

\bibitem{ma2008enhanced}
Ma Y, Hao Q, Poudel B, Lan Y, Yu B, Wang D, Chen G and Ren Z 2008 {\em Nano Lett.\/} {\bf 8} 2580--2584

\bibitem{xie2009unique}
Xie W, Tang X, Yan Y, Zhang Q and Tritt T~M 2009 {\em Appl. Phys. Lett.\/} {\bf 94}

\bibitem{xie2010identifying}
Xie W, He J, Kang H~J, Tang X, Zhu S, Laver M, Wang S, Copley J~R, Brown C~M, Zhang Q {\em et~al.\/} 2010 {\em Nano Lett.\/} {\bf 10} 3283--3289

\bibitem{yan2013thermoelectric}
Yan X, Liu W, Chen S, Wang H, Zhang Q, Chen G and Ren Z 2013 {\em Adv. Energy Mater.\/} {\bf 3} 1195--1200

\bibitem{mahan1996best}
Mahan G and Sofo J 1996 {\em Proc. Natl. Acad. Sci. U.S.A.\/} {\bf 93} 7436--7439

\bibitem{mao2015high}
Mao J, Wang Y, Kim H~S, Liu Z, Saparamadu U, Tian F, Dahal K, Sun J, Chen S, Liu W {\em et~al.\/} 2015 {\em Nano Energy\/} {\bf 17} 279--289

\bibitem{biswas2012high}
Biswas K, He J, Blum I~D, Wu C~I, Hogan T~P, Seidman D~N, Dravid V~P and Kanatzidis M~G 2012 {\em Nature\/} {\bf 489} 414--418

\bibitem{ren2017enhancing}
Ren G~K, Wang S~Y, Zhu Y~C, Ventura K~J, Tan X, Xu W, Lin Y~H, Yang J and Nan C~W 2017 {\em Energy Environ. Sci.\/} {\bf 10} 1590--1599

\bibitem{Pei2011Ag2Te}
Pei Y, Heinz N~A and Snyder G~J 2011 {\em J. Mater. Chem.\/} {\bf 21} 18256--18260

\bibitem{bilc2004resonant}
Bilc D, Mahanti S, Quarez E, Hsu K~F, Pcionek R and Kanatzidis M 2004 {\em Phys. Rev. Lett.\/} {\bf 93} 146403

\bibitem{Heremans2012bulk}
Heremans J~P, Wiendlocha B and Chamoire A~M 2012 {\em Energy Environ. Sci.\/} {\bf 5} 5510--5530 ISSN 1754-5692

\bibitem{pei2012low}
Pei Y, LaLonde A~D, Wang H and Snyder G~J 2012 {\em Energy Environ. Sci.\/} {\bf 5} 7963--7969

\bibitem{usui2017enhanced}
Usui H and Kuroki K 2017 {\em J. Appl. Phys\/} {\bf 121}

\bibitem{usui2013large}
Usui H, Suzuki K, Kuroki K, Nakano S, Kudo K and Nohara M 2013 {\em Phys. Rev. B\/} {\bf 88} 075140

\bibitem{kuroki2007pudding}
Kuroki K and Arita R 2007 {\em J. Phys. Soc. Jpn.\/} {\bf 76} 083707--083707

\bibitem{isaacs2019remarkable}
Isaacs E~B and Wolverton C 2019 {\em Phys. Rev. Mater\/} {\bf 3} 015403

\bibitem{wei2020strain}
Wei S, Wang C, Fan S and Gao G 2020 {\em J. Appl. Phys\/} {\bf 127}

\bibitem{xia2019leveraging}
Xia Y, Park J, Ozoli{\c{n}}{\v{s}} V and Wolverton C 2019 {\em Phys. Rev. B\/} {\bf 100} 201401

\bibitem{xia2019high}
Xia Y, Park J, Zhou F and Ozoli{\c{n}}{\v{s}} V 2019 {\em Phys. Rev. Appl.\/} {\bf 11} 024017

\bibitem{adhidewata2022thermoelectric}
Adhidewata J~M, Nugraha A~R, Hasdeo E~H, Estell{\'e} P and Gunara B~E 2022 {\em Mater. Today Commun.\/} {\bf 31} 103737

\bibitem{adhidewata2022thermoelectrics}
Adhidewata J~M, Nugraha A~R, Hasdeo E~H and Gunara B~E 2022 {\em Indones. J. Appl. Phys.\/} {\bf 33} 51--57

\bibitem{hasdeo2019optimal}
Hasdeo E~H, Krisna L~P~A, Hanna M~Y, Gunara B~E, Hung N~T and Nugraha A~R~T 2019 {\em J. Appl. Phys\/} {\bf 126} 035109 ISSN 0021-8979

\bibitem{Shao2020}
Shao Y, Rudenko A~N, Hu J, Sun Z, Zhu Y, Moon S, Millis A~J, Yuan S, Lichtenstein A~I, Smirnov D, Mao Z~Q, Katsnelson M~I and Basov D~N 2020 {\em Nat. Phys.\/} {\bf 16} 636--641 ISSN 1745-2481

\bibitem{Rudenko2018}
Rudenko A, Stepanov E, Lichtenstein A and Katsnelson M 2018 {\em Phys. Rev. Lett.\/} {\bf 120} 216401

\bibitem{Singha2017}
Singha R, Pariari A~K, Satpati B and Mandal P 2017 {\em Proc. Natl. Acad. Sci. U.S.A.\/} {\bf 114} 2468--2473

\bibitem{ali2016butterfly}
Ali M~N, Schoop L~M, Garg C, Lippmann J~M, Lara E, Lotsch B and Parkin S~S 2016 {\em Sci. Adv.\/} {\bf 2} e1601742

\bibitem{Guan2017}
Guan S, Yu Z~M, Liu Y, Liu G~B, Dong L, Lu Y, Yao Y and Yang S~A 2017 {\em npj Quantum Mater.\/} {\bf 2} 23 ISSN 2397-4648 \urlprefix\url{https://doi.org/10.1038/s41535-017-0026-7}

\bibitem{zhang2017topological}
Zhang X, Jin L, Dai X and Liu G 2017 {\em J. Phys. Chem. Lett.\/} {\bf 8} 4814--4819

\bibitem{wang2020unique}
Wang X, Ding G, Khandy S~A, Cheng Z, Zhang G, Wang X~L and Chen H 2020 {\em Nanoscale\/} {\bf 12} 16910--16916

\bibitem{pan2021thermoelectric}
Pan Y, Fan F~R, Hong X, He B, Le C, Schnelle W, He Y, Imasato K, Borrmann H, Hess C {\em et~al.\/} 2021 {\em Adv. Energy Mater.\/} {\bf 33} 2003168

\bibitem{rudderham2021ab}
Rudderham C and Maassen J 2021 {\em Phys. Rev. B\/} {\bf 103} 165406

\bibitem{bahk2016minority}
Bahk J~H and Shakouri A 2016 {\em Phys. Rev. B\/} {\bf 93} 165209

\bibitem{androulakis2010thermoelectric}
Androulakis J, Todorov I, Chung D~Y, Ballikaya S, Wang G, Uher C and Kanatzidis M 2010 {\em Phys. Rev. B\/} {\bf 82} 115209

\bibitem{gayner2016boost}
Gayner C, Sharma R, Das M~K and Kar K~K 2016 {\em J. Appl. Phys\/} {\bf 120}

\bibitem{hung2022enhanced}
Hung N~T, Adhidewata J~M, Nugraha A~R and Saito R 2022 {\em Phys. Rev. B\/} {\bf 105} 115142

\bibitem{goldsmid2010introduction}
Goldsmid H~J {\em et~al.\/} 2010 {\em Introduction to thermoelectricity\/} vol 121 (Springer)

\bibitem{ashcroft1976nd}
Ashcroft N~W 1976 {\em Saunders College, Philadelphia\/} {\bf 120}

\bibitem{markov2018semi}
Markov M, Hu X, Liu H~C, Liu N, Poon S~J, Esfarjani K and Zebarjadi M 2018 {\em Sci. Rep.\/} {\bf 8} 9876

\bibitem{chasapis2015understanding}
Chasapis T~C, Lee Y, Hatzikraniotis E, Paraskevopoulos K~M, Chi H, Uher C and Kanatzidis M~G 2015 {\em Phys. Rev. B\/} {\bf 91} 085207

\bibitem{neto2009electronic}
Neto A~C, Guinea F, Peres N~M, Novoselov K~S and Geim A~K 2009 {\em Rev. Mod. Phys.\/} {\bf 81} 109

\bibitem{ariel2012energy}
Ariel V 2012 {\em arXiv preprint arXiv:1207.4282\/}

\bibitem{wickramaratne2015electronic}
Wickramaratne D, Zahid F and Lake R~K 2015 {\em J. Appl. Phys\/} {\bf 118}

\bibitem{giannozzi2009quantum}
Giannozzi P, Baroni S, Bonini N, Calandra M, Car R, Cavazzoni C, Ceresoli D, Chiarotti G~L, Cococcioni M, Dabo I {\em et~al.\/} 2009 {\em J. Phys. Condens. Matter\/} {\bf 21} 395502

\bibitem{curtarolo2012aflowlib}
Curtarolo S, Setyawan W, Wang S, Xue J, Yang K, Taylor R~H, Nelson L~J, Hart G~L, Sanvito S, Buongiorno-Nardelli M {\em et~al.\/} 2012 {\em Comput. Mater. Sci.\/} {\bf 58} 227--235

\bibitem{perdew1992atoms}
Perdew J~P, Chevary J~A, Vosko S~H, Jackson K~A, Pederson M~R, Singh D~J and Fiolhais C 1992 {\em Phys. Rev. B\/} {\bf 46} 6671

\bibitem{madsen2018boltztrap2}
Madsen G~K, Carrete J and Verstraete M~J 2018 {\em Comput. Phys. Commun.\/} {\bf 231} 140--145

\bibitem{virtanen2020scipy}
Virtanen P, Gommers R, Oliphant T~E, Haberland M, Reddy T, Cournapeau D, Burovski E, Peterson P, Weckesser W, Bright J {\em et~al.\/} 2020 {\em Nat. Methods\/} {\bf 17} 261--272

\bibitem{GithubNorman}
Python codes to obtain all calculation results and parameters from this paper are available at https://github.com/Normanthen/NLS-Thermoelectrics , where we also provide additional calculations for variations of $v_{F}$ and $m$ beyond what is discussed in the main text.

\bibitem{ali2019thermoelectric}
Ali A, Rahman A~U and Rahman G 2019 {\em Phys. B: Condens. Matter\/} {\bf 565} 18--24

\bibitem{xu2020centrosymmetric}
Xu L, Zhang X, Meng W, He T, Liu Y, Dai X, Zhang Y and Liu G 2020 {\em  J. Mater. Chem. C\/} {\bf 8} 14109--14116

\bibitem{soleimani2020review}
Soleimani Z, Zoras S, Ceranic B, Shahzad~S, and Cui Y 2020 {\em Sustain. Energy Technol. Assess.} {\bf 37} 100604

\end{thebibliography}

\providecommand{\newblock}{}

\end{document}